\documentclass[12pt]{iopart}
\usepackage{graphicx}
\usepackage{subfigure}

\begin{document}

\title{Classical to quantum transition of a driven nonlinear
  nanomechanical resonator}%
\author{Itamar Katz and Ron Lifshitz}%
\address{Raymond and Beverly Sackler School of Physics \& Astronomy,
  Tel Aviv University, Tel Aviv 69978, Israel}%
\ead{ronlif@tau.ac.il}%
\author{Alex Retzker}%
\address{Institute for Mathematical Sciences, Imperial College, London
  SW7 2PE, U.K.}%
\author{Raphael Straub}%
\address{Department of Physics, University of Konstanz, D-78457
  Konstanz, Germany}%

\begin{abstract}
  Much experimental effort is invested these days in fabricating
  nanoelectromechanical systems (NEMS) that are sufficiently small,
  cold, and clean, so as to approach quantum mechanical behavior as
  their typical quantum energy scale $\hbar\Omega$ becomes comparable
  to that of the ambient thermal energy $k_{B}T$. Such systems will
  hopefully enable one to observe the quantum behavior of human-made
  objects, and test some of the basic principles of quantum mechanics.
  Here we expand and elaborate on our recent suggestion
  [\textit{Phys.\ Rev.\ Lett.} {\bf 99} (2007) 040404] to exploit the
  nonlinear nature of a nanoresonator in order to observe its
  transition into the quantum regime. We study this transition for an
  isolated resonator, as well as one that is coupled to a heat bath at
  either zero or finite temperature. We argue that by exploiting
  nonlinearities, quantum dynamics can be probed using technology that
  is almost within reach.  Numerical solutions of the equations of
  motion display the first quantum corrections to classical dynamics
  that appear as the classical-to-quantum transition occurs. This
  provides practical signatures to look for in future experiments with
  NEMS resonators.
\end{abstract}

\maketitle

\section{Motivation: Mechanical Systems at the Quantum Limit}

This special focus issue of the \emph{New Journal of Physics} is
motivated by the growing interest in the study of ``Mechanical Systems
at the Quantum Limit''. Nanoelectromechanical systems
(NEMS)~\cite{craighead00,roukes01_02,roukes01_01,C03,EkinciRoukes05}
offer one of the most natural playgrounds for such a
study~\cite{cho03,blencowe04,schwab_roukes05}. With recent experiments
coming within less than an order of magnitude from the ability to
observe quantum zero-point motion~\cite{knobel03,LaHaye04,Naik06},
ideas about the quantum-to-classical transition
(QCT)~\cite{Penrose96,Leggett99,Leggett02} may soon become
experimentally accessible, more than 70 years after Schr\"{o}dinger
described his famous cat paradox~\cite{cat}. As nanomechanical
resonators become smaller, their masses decrease and natural
frequencies $\Omega$ increase---exceeding 1GHz in recent
experiments~\cite{HZMR03,cleland04}. For such frequencies it is
sufficient to cool down to temperatures on the order of $50mK$ for the
quantum energy $\hbar\Omega$ to be comparable to the thermal energy
$k_{B}T$.  Cooling well below such temperatures should allow one to
observe truly quantum mechanical phenomena, such as resonances,
oscillator number states, superpositions, and
entanglement~\cite{peano04,Carr01,Armour02,santamore04,Eisert04}, at
least for macroscopic objects that are sufficiently isolated from
their environment.

Here we expand and elaborate on our recent
suggestion~\cite{myPRLarticle} to exploit the nonlinear nature of NEMS
resonators~\cite{LCReview} in order to study the transition into the
quantum domain. Nonlinear behavior of NEMS (and MEMS) resonators is
frequently observed in experiments~\cite{craighead00,turner98,BR00,%
  BR02,Blick02_2,yu02,aldridge05,ina05,ina06,kozinsky07}, offers
interesting theoretical challenges~\cite{lifshitz03,bromberg06,%
  cross04,cross06}, and can also be exploited for applications~\cite{%
  rugar91,carr00,lupascu06,milburn_schwab08}. Here we study a driven
nonlinear nanomechanical resonator, also known as the Duffing
resonator. We choose the system parameters such that the classical
Duffing resonator is in a regime in which, in the presence of
dissipation, it can oscillate at one of two different amplitudes, or
one of two dynamical steady-states of motion, depending on its initial
conditions.  This dependence on initial conditions has recently been
mapped-out by Kozinsky \emph{et al.}~\cite{kozinsky07} in an
experiment with a nanoresonator.  Thermally-induced switching between
the two dynamical steady-states of motion has also been observed in a
nanoresonator experiment by Aldridge and Cleland~\cite{aldridge05}.

Naturally, the question arises whether quantum-mechanically the
resonator can be driven into a superposition of the two possible
steady-states of motion, or at least tunnel or switch between them at
temperatures that are sufficiently low for thermal switching to be
suppressed, but not necessarily so low to satisfy the constraint
$k_{B}T\ll\hbar\Omega$ for observing full-fledged quantum mechanics.
To answer this question we perform two separate calculations on the
same Duffing resonator, viewing it once as a quantum-mechanical object
and once as a classical object.  With everything else being equal,
this allows us to contrast the dynamics of a classical resonator with
that of its quantum clone. We can then search for a regime in which
quantum dynamics just begins to deviate away from classical dynamics,
providing us with practical signatures that can be detected
experimentally.

We calculate the dynamics numerically. We start each quantum
calculation with a coherent state, which is a minimal wave-packet
centered about some point in phase space. We start the corresponding
classical calculation with an ensemble of initial
conditions---typically of $N=10^{4}$ points---drawn from a Gaussian
distribution in phase space that is identical to the initial
quantum-mechanical probability density. We perform these calculations
for three qualitatively different situations: An isolated resonator
with no coupling to the environment; a resonator coupled to a heat
bath at temperature $T_{env}=0$; and a resonator coupled to a heat
bath at a finite temperature $T_{env}>0$. In all cases we display the
calculated quantum dynamics in phase space using the quantum Wigner
function
\begin{equation}\label{wigner}
W(x,p,t) = {\frac{1}{\pi\hbar}} \int_\infty^\infty dx'
e^{-{2i\over\hbar}px'} \langle x+x'|\rho(t)|x-x'\rangle,
\end{equation}
where $\rho(t)$ is the usual density operator, and compare it with the
time-evolution of the classical phase-space density. We remind the
reader that the Wigner function is not a true probability distribution
as it may possess negative values, particularly when the quantum state
has no classical analog. Nevertheless, it reduces to the quantum
probability $P(x,t)$ of observing the system at position $x$ at time
$t$ upon integration over $p$, and {\it vice versa.}\footnote{There
  exist other possibilities for describing quantum dynamics in phase
  space, see for
  example~\cite{gardiner_zoller04,walls_milburn94,sergi-2007}, but we
  find the Wigner function to be among the simplest.} 

The paper is organized as follows. Section~\ref{QCT} gives a general
discussion of the transition from quantum to classical dynamics, and
the corresponding limit of $\hbar\to0$.  Section~\ref{closed system
  dynamics} describes the classical and quantum mechanical equations
of motion for an isolated resonator, and section~\ref{closed system
  results} discusses the results of our calculation for the isolated
case. In Section~\ref{open system dynamics} we describe the
dynamics of a resonator that is coupled to a heat bath, and in
Sections~\ref{open system T zero results} and~\ref{open system T
  nonzero results} we present the results of our calculation for a
resonator, coupled to a zero-temperature and a finite-temperature bath
respectively. We summarize and discuss the significance of our results
in Section \ref{conclusion}, concluding with some ideas for future
directions.

\goodbreak
\section{The Quantum to Classical Transition (QCT)}\label{QCT}

Everyday macroscopic objects behave according to classical physics,
expressed in terms of Newton's or Hamilton's equations of motion.
Since these objects are composed of atoms and molecules, one expects
this classical behavior to emerge from quantum dynamics under a
certain limit of high temperature, large masses, and high energies, all
characterizing the macroscopic world. It is commonly accepted that the
transition between the quantum and the classical descriptions occurs
by letting $\hbar$ go to zero.  Of course $\hbar$ is actually a
non-zero physical constant with the dimensionality of action, so
whenever one says that ``$\hbar$ goes to zero'' one actually means
that some dimensionless combination of $\hbar$ with a physical
quantity characterizing the system---like its ratio to the classical
action $\hbar/S$---goes to zero. 

One can see in a number of different ways that classical dynamics
should be obtained from the quantum description by letting
$\hbar\rightarrow0$.  If one expresses the quantum mechanical wave
function as $\psi=\sqrt{\rho}\exp[iS/\hbar]$, then upon letting
$\hbar\rightarrow0$ the Schr\"odinger equation reduces to the
classical Hamilton-Jacoby equation for $S$~\cite[Section 6.4]{H&F}.
The Feynman path integral approach to quantum mechanics, in which one
sums over all paths between two points in space and time, reduces to
Hamilton's principle of classical mechanics, identifying the classical
path for which the action satisfies $\delta S=0$~\cite{feynman_hibbs}.
Finally, one can look at the equation of motion for the quantum Wigner
function~(\ref{wigner}).  This equation can be obtained using the von
Neumann equation for the density operator $\rho(t)$,
\begin{equation}\label{VN}
\dot{\rho} = \frac{1}{i\hbar}[H_{sys},\rho],
\end{equation}
where $H_{sys}$ is the Hamiltonian of the system. For a general
potential $U(x)$, which can be expanded in a Taylor series, one gets
the quantum Liouville equation~\cite{schleich01},
\begin{equation}\label{quantum liouville}
  \fl\left(\partial_{t}+\frac{p}{m}\partial_{x} -
  \frac{dU(x)}{dx}\partial_{p}\right)W(x,p,t) =
  \sum_{n=1}^{\infty}\frac{(-1)^{n}(\hbar/2)^{2n}}{(2n+1)!}
  \partial_{x}^{2n+1}U(x)\partial_{p}^{2n+1}W(x,p,t).    
\end{equation}
If one formally sets $\hbar=0$, and unless the derivatives of the
Wigner function on the right-hand side become singular, the right-hand
side is equal to zero and one recovers the classical Liouville
equation for the distribution in phase space.\footnote{This is also
  true for a potential containing terms up to quadratic in the
  position, even for finite $\hbar$. Therefore, the quantum and
  classical dynamics of a harmonic oscillator are identical in the
  sense of the time evolution of phase-space distributions and the
  expectation values which can be calculated from them.}

\label{singularity of hbar}
It turns out that this naive approach does not work in general, and
that one often encounters non-analyticities in quantum mechanics as
one takes the limit $\hbar\rightarrow0$~\cite{berry89}. For an
isolated system, the classical Liouville and master equations (see
Section 3) violate unitarity, as well as the quantum restriction of
the density matrix to be semi-positive definite. Thus, the limit
$\hbar\rightarrow0$ cannot be connected smoothly to
$\hbar=0$.\footnote{A simple argument is given by Habib \textit{et
    al.}~\cite{habibPRL} to explain this violation. An interesting
  approach to overcome this difficulty is presented, for example, by
  Gat~\cite{gat07}.} Therefore, as long as the dynamics considered is
of an isolated system, the evolution of a quantum observable will in
general agree with the corresponding classical average only up to a
finite time, called the \emph{Ehrenfest time}, $t_{E}$.  For non
chaotic systems, it is expected that $t_{E}$ will have some power law
dependence on $\hbar$, $t_{E}\sim\hbar^{-\delta}$, though the exact
value of $\delta$ may depend on the
model~\cite{berry78,habib98,cametti02}.  Nevertheless, we observe here
that it is difficult strictly to define the time scale at which a
separation between classical and quantum dynamics occurs, mainly
because this time scale seems to depend on the observable one chooses
to measure.  Generally, it seems that there is no single time scale,
as described by Oliveira \textit{et al.}~\cite{oliveira03,oliveira06}.

The contemporary approach to understanding the classical limit of
quantum mechanics---stemming from early work of Feynman and
Vernon~\cite{Feynman_Vernon63} and Caldeira and
Leggett~\cite{caldeira_leggett83}---is to look at a more realistic
model, in which the system of interest is coupled in some way to the
external world, generically referred to as `an environment', whether
it is another system or a measuring device~\cite{gardiner_zoller04,%
  walls_milburn94,schlosshauer07}.  The interaction with the
environment must be taken into account, and one distinguishes between
two types of systems, depending on whether a measurement is performed
on the environment or not~\cite{MCWF2,greenbaum05,habibLANL}. If the
environment is not observed, then the appropriate description of the
system is in terms of the reduced density operator, obtained by
tracing the full density matrix over the variables of the environment.
Time evolution is then given by a master equation, and the so called
\textit{weak form of QCT} is obtained by comparing quantum and
classical distributions.  This is what we do here, while noting in
agreement with Habib \emph{et al.}~\cite{habib98} that the role of the
environment is actually two-fold. On one hand, it destroys
interference patterns with negative values in the Wigner function that
are incompatible with a classical probability density. On the other
hand, it causes the classical fine structure to smear over a thermal
scale, producing a classical distribution without the infinitely-small
fine structure, which is incompatible with quantum mechanics owing to
the finite value of $\hbar$ and the uncertainty principle.

If the environment is measured, the reduced density matrix depends on
the outcome of the measurements, and the evolution is said to be
conditioned on the observation results, as described for example by
Habib \textit{et al.}~\cite{habibLANL}. This type of evolution can
yield effectively classical trajectories, and is called \textit{the
  strong form of QCT}. We shall analyze the latter case, which
requires more advanced tools of quantum measurement theory, in a
future publication.

\section{Isolated System -- Method of Calculation}
\label{closed system dynamics}

We consider a nonlinear resonator, such as a doubly-clamped
nanomechanical beam~\cite{ina05,ina06},
nanowire~\cite{husain03,feng07}, or
nanotube~\cite{yaish04,vanderzant06}, vibrating in its fundamental
flexural mode, and thus treated as a single degree of freedom. The
resonator is driven, in any of the standard NEMS
techniques~\cite{C03,EkinciRoukes05}, by an external periodic force
and is either isolated from its environment---a case treated in this
section---or coupled to the environment in the form of a heat bath at
a temperature $T_{env}$---a case that will be treated later. The
Hamiltonian for such a driven nonlinear resonator, known as the
Duffing resonator, is
\begin{equation}\label{Classical Hamiltonian}
  \tilde{H}_{sys} = \frac{1}{2m}\tilde{p}^{2} +
  \frac{m\Omega^2}{2}\tilde{x}^{2} + \frac{\varepsilon}{4}\tilde{x}^{4} -
  \tilde{x}\tilde{F}\cos\tilde{\omega}\tilde{t},   
\end{equation}
where $m$ is the effective mass of the resonator, $\Omega$ its normal
frequency, $\varepsilon$ the strength of nonlinearity, $\tilde{F}$ the
driving amplitude, and $\tilde{\omega}$ the driving frequency.
Throughout the discussion here, and with no loss of generality, we
consider the case of a stiffening nonlinearity, with $\varepsilon>0$.

We change from the physical variables, denoted by tildes, to
dimensionless variables without tildes, by measuring time in units of
$1/\Omega$, mass in units of $m$, and length in units of
$\sqrt{m\Omega^2/\varepsilon}$. The dimensionless Hamiltonian
$H_{sys}$ is then measured in units of the characteristic energy
$m^2\Omega^4/\varepsilon$,
\begin{equation}
  H_{sys} = \frac{1}{2}p^2 + \frac{1}{2}x^2 + \frac{1}{4}x^4 -
  xF\cos\omega t, 
\end{equation}
where $x$, $p$, and $t$ are now measured in the dimensionless units,
and $F=\tilde{F}\sqrt{\varepsilon}(m\Omega^2)^{-\frac{3}{2}}$ and
$\omega=\tilde{\omega}/\Omega$ are the dimensionless driving amplitude
and frequency, respectively. The degree to which we approach the
quantum domain is indicated by an increasing effective value of
$\hbar$ as compared with a measure of the classical action $S$ of the
system, both stated in terms of the scaled units
$m^2\Omega^3/\varepsilon$.  For the case of transverse vibrations of a
doubly-clamped beam, as shown by Lifshitz and Cross~\cite{LCReview},
the nonlinear coefficient $\varepsilon$ is given by $m\Omega^2/d^2$ to
within a numeral factor of order unity, where $d$ is the width, or
diameter, of the beam.  Thus, the scaled units of action become
$m\Omega d^2$. We shall reinterpret these values later in terms of
real experimental masses, frequencies, and widths.

\subsection{Classical dynamics -- Hamilton and Liouville equations}
\label{closed classical dynamics}

Hamilton's equations give the equations of motion for $x$ and $p$,
\begin{eqnarray}
\dot{x} &=& p\\
\dot{p} &=& -x-x^{3}+F\cos\omega t.
\label{closed classical eom}
\end{eqnarray}
For the isolated resonator, the classical Liouville equation for
the phase-space distribution function $f(x,p,t)$ is
\begin{equation}
\frac{\partial f}{\partial t} = -\{f,H_{sys}\},
\end{equation}
where $\{\cdot,\cdot\}$ denotes the Poisson brackets. Upon
substituting the Hamiltonian~(\ref{Classical Hamiltonian}) one gets
(denoting $\partial_{x}\equiv{\partial}/{\partial x}$ etc.)
\begin{equation}\label{classical liouville}
\partial_{t} f=-\left[p\partial_{x}+\left(-x-x^3+F\cos\omega
t\right)\partial_{p}\right]f,
\end{equation}
where $x$ and $p$ are treated as independent variables. The initial
state $f(x,p,t=0)$ is taken to be a Gaussian distribution centered at
some $[x_{0},p_{0}]$ and having standard deviations of $\Delta
x=\sqrt{\hbar/2}$ and $\Delta p=\sqrt{\hbar/2}$, corresponding to an
initial quantum coherent state with minimum uncertainty $\Delta
x\Delta p=\hbar/2$.

We can examine the case of a linear resonator simply by omitting the
cubic term in~(\ref{closed classical eom}). In this case the result is
a set of linear ODE's that can be solved analytically. For the
nonlinear case we integrate the equation of motion (\ref{closed
  classical eom}) numerically. The solution enables us to calculate
ensemble averages and make plots of the density in phase space.
Instead of keeping track of all the trajectories as a function of
time, we calculate a histogram of the probability density $f(x,p,t)$.
We do so by dividing phase space using a grid, where each entry of the
grid counts the number of trajectories passing through a particular
square in phase space at any given time. This grid is then used to
calculate the classical averages. The discrete nature of the grid and
the finite number of initial conditions (typically 10$^4$) is the
reason for the apparent differences between the initial classical and
quantum states shown in the Figures below.

\subsection{Quantum dynamics -- The Schr\"{o}dinger equation}
\label{closed quantum dynamics}

The quantum Hamiltonian operator of the driven Duffing resonator is
the same as (\ref{Classical Hamiltonian}),
\begin{equation}\label{Quantum Hamiltonian}
  H_{sys} = \hbar\left(a^{\dagger}a + \frac{1}{2}\right) +
  \frac{\hbar^2}{16}\left(a^{\dagger}+a\right)^{4} - 
  \sqrt{\frac{\hbar}{2}}\left(a^{\dagger}+a\right)F\cos\omega t, 
\end{equation}
where we have used the ordinary definition of the ladder operators
\numparts
\begin{eqnarray}\label{eq:ladder}
a &=& \frac{1}{\sqrt{2\hbar}}(x +ip),\\
a^{\dag} &=&
\frac{1}{\sqrt{2\hbar}}(x-ip).
\end{eqnarray}
\endnumparts We simply solve the corresponding Schr\"{o}dinger
equation numerically by advancing the wave function in discrete time
steps $\Delta t$, using the interaction-picture time-evolution
operator.  We do so by expanding the wave function in a truncated
energy basis $|n\rangle$, with $n\leq N$. The value of $N$, which is
typically $\approx50$, is determined by the requirement that the
high-energy basis states are never significantly occupied. We use the
wave function at time $t$ to calculate and then plot the corresponding
Wigner function.

\section{Isolated System -- Results}
\label{closed system results}

\begin{figure}[bt]
  \begin{center}
    \subfigure[Initial coherent state, $t=0$]{
      \includegraphics[clip=true, viewport=0.6cm 2.2cm 13.5cm
      8.5cm,width=0.45\textwidth]{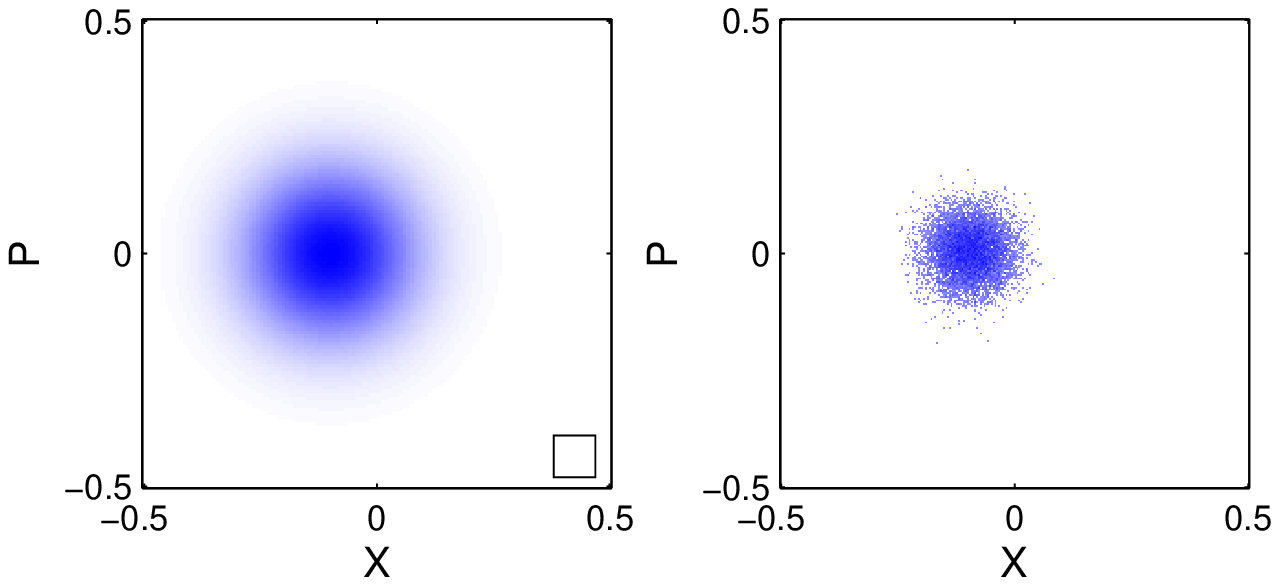}}
    \subfigure[$t=30T$]{
      \includegraphics[clip=true, viewport=0.6cm 2.2cm 13.5cm
      8.5cm,width=0.45\textwidth]{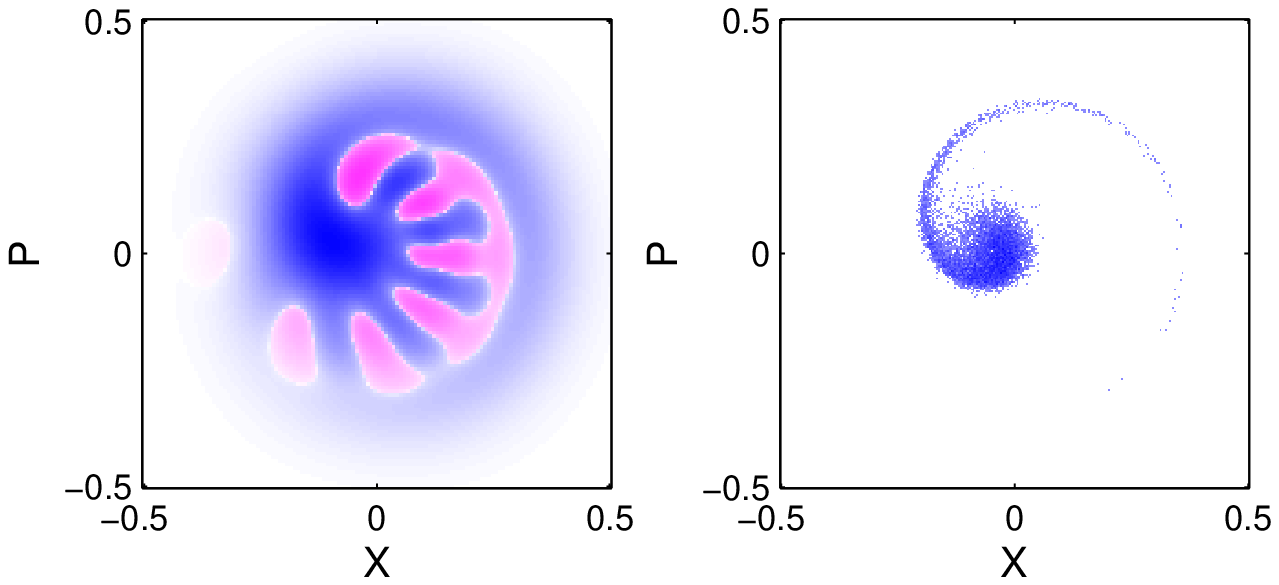}}
    \subfigure[$t=156T$]{
      \includegraphics[clip=true, viewport=0.6cm 2.2cm 13.5cm
      8.5cm,width=0.45\textwidth]{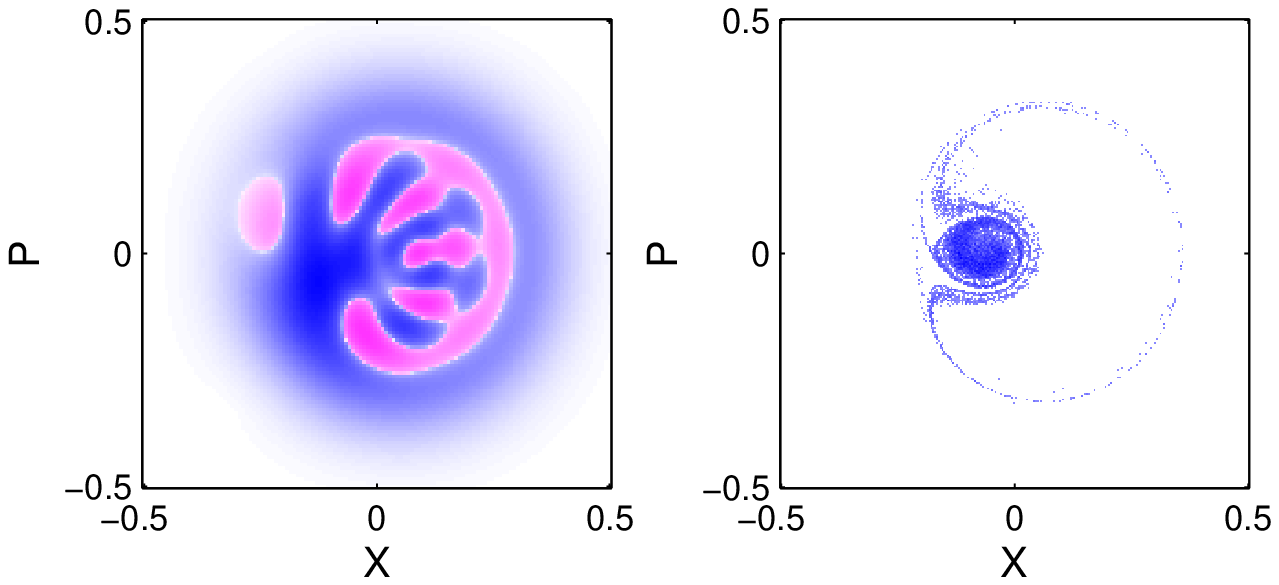}}
    \subfigure[$t=300T$]{
      \includegraphics[clip=true, viewport=0.6cm 2.2cm 13.5cm
      8.5cm,width=0.45\textwidth]{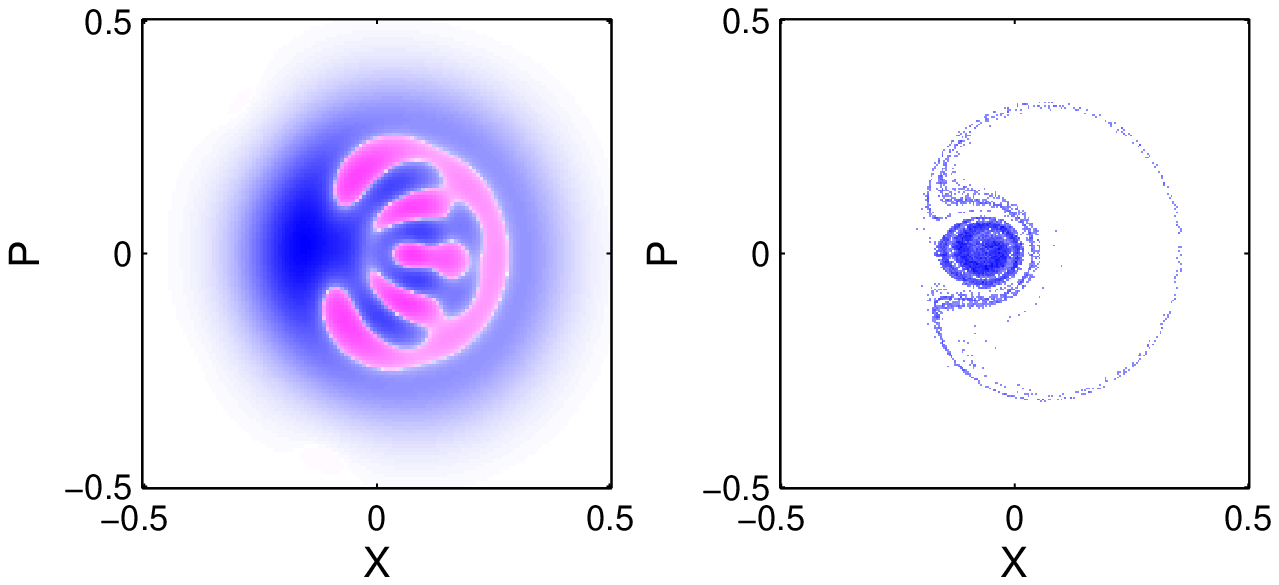}}
    \end{center}
    \caption[Isolated Duffing resonator with $\hbar=0.008$. Wigner
    function and classical distribution.]  {Isolated Duffing resonator
      with $\hbar=0.008$. Wigner function (left) and classical
      distribution (right) of the initial minimal Gaussian wave packet
      and its evolution at three later times, measured in terms of the
      drive period $T=2\pi/\omega$. A square of area $\hbar$ is shown
      at the bottom right corner of the initial state plot. The
      functions are scaled by $f\rightarrow f^{1/4}$ for better color
      contrast.% A link for the full Poincar\`{e} section time
    % evolution is given in appendix C(4).
    }
    \label{nonlinear close 0.2}
\end{figure}

All results in this section are given for a choice of parameters
$F=0.01$ and $\omega=1.018$, which ensures that the resonator is in
the bistability region.  The results are shown in
Figure~\ref{nonlinear close 0.2} by plotting the Wigner function next
to the sampled classical phase-space density, for the initial minimal
Gaussian wave packet and its later evolution at three different times,
measured in periods of the drive $T=2\pi/\omega$. We note that the
apparent discreteness of the classical distributions, as opposed to
the smooth quantum Wigner functions, stems from the fact that the
former are generated from a histogram on a discrete grid.  The
densities are raised to the power $1/4$ for better color contrast,
where blue denotes positive values, and red denotes negative values.
A square of area $\hbar$ is shown at the bottom right corner of the
initial state plot to indicate the scale of $\hbar$.

We see that for short times the positive (blue) outline of the Wigner
function resembles that of the classical distribution (Figure
\ref{nonlinear close 0.2}b). In addition, a strong interference
pattern is evident within this outline. In all the snapshots it is
evident that the Wigner function has a strong positive density where
the classical density is also large. Nevertheless, two differences
between the quantum and classical distributions are evident, namely,
the strong interference pattern which exists in the Wigner function
even in the steady state, and the infinitely fine structure which
develops in the classical distribution. These differences demonstrate that
taking the limit of $\hbar\to 0$ does not give rise to the emergence
of  classical dynamics from the quantum description, as indicated earlier
in Section~\ref{singularity of hbar}.

\begin{figure}[tb]
  \centering
  \subfigure[Original distributions]{
    \includegraphics[clip=true, viewport=0.8cm 2.2cm 13.5cm
    9cm,width=0.45\textwidth]{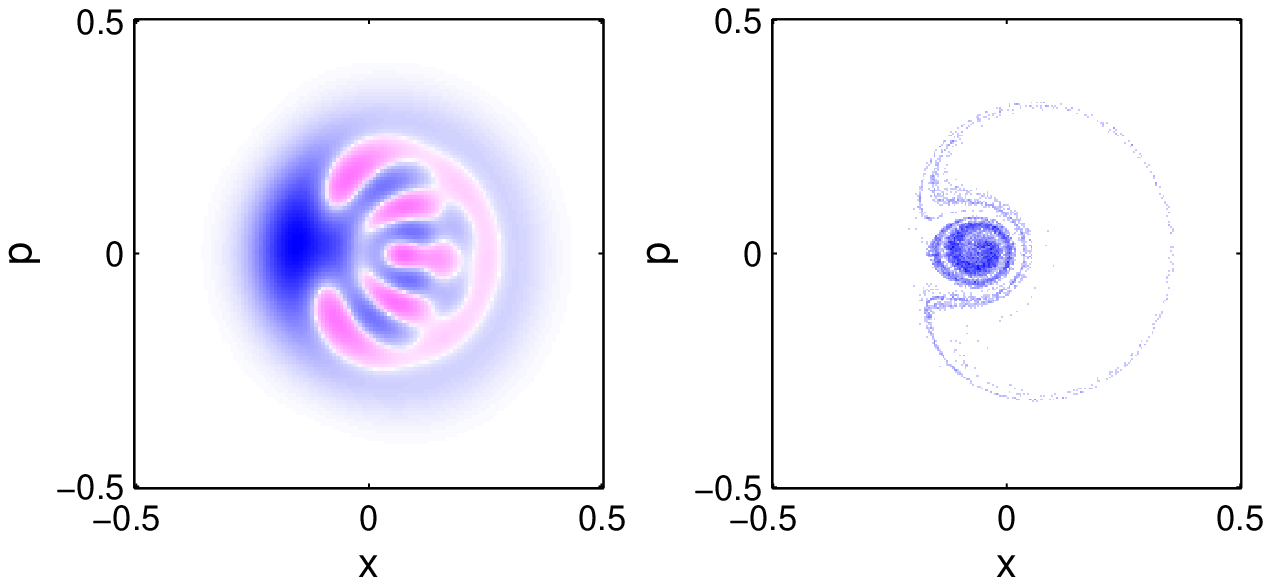}}
  \subfigure[Averaged distributions]{
    \includegraphics[clip=true, viewport=0.8cm 2.2cm 13.5cm
    9cm,width=0.45\textwidth]{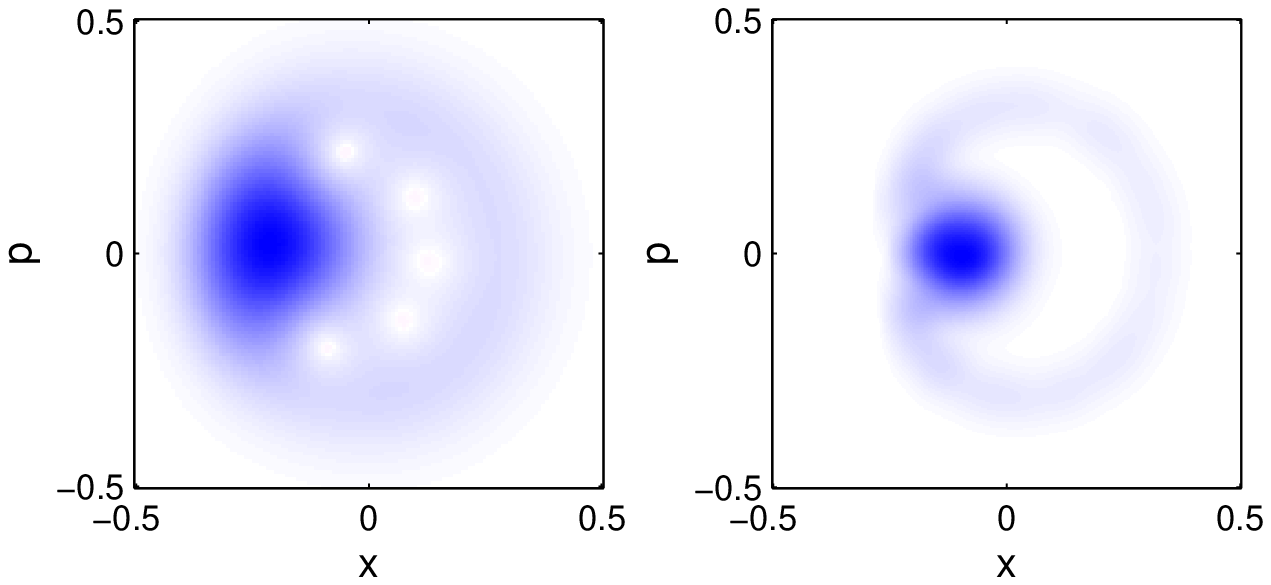}}
  \caption[Averaged distributions for the isolated Duffing
  resonator]{Isolated Duffing resonator with $\hbar=0.008$. (a)
    Original Wigner function and classical distribution (left and
    right, respectively), shown after 300 periods of the drive, as
    obtained from the calculations shown in Fig.~\ref{nonlinear close
      0.2}. (b) The same distributions averaged using a Gaussian
    kernel of area $2\hbar$.  The functions are scaled to the power of
    $1/2$.} \label{averaged wigner}
\end{figure}

We can, however, use the approach proposed by Berry~\cite{berry89},
where the quantum probability density is averaged over $x$ and $p$ to
account for the limited precision of a typical measuring device. If we
do the same here, and replace the value of the Wigner function
$W(x,p)$ with its averaged value over a small area of size $\delta
x\delta p$ around the point $(x,p)$, the negative and positive parts
of the interference pattern cancel each other out. If we perform the
same averaging also for the classical distribution, the classical fine
structure is smeared, bringing the two distributions even closer in
appearance, as shown in Figure~\ref{averaged wigner}.

\begin{figure}[bt]
    \begin{center}
    \subfigure[$n=1, \hbar=0.032$]{
    \includegraphics[height=4cm,width=0.4\textwidth]{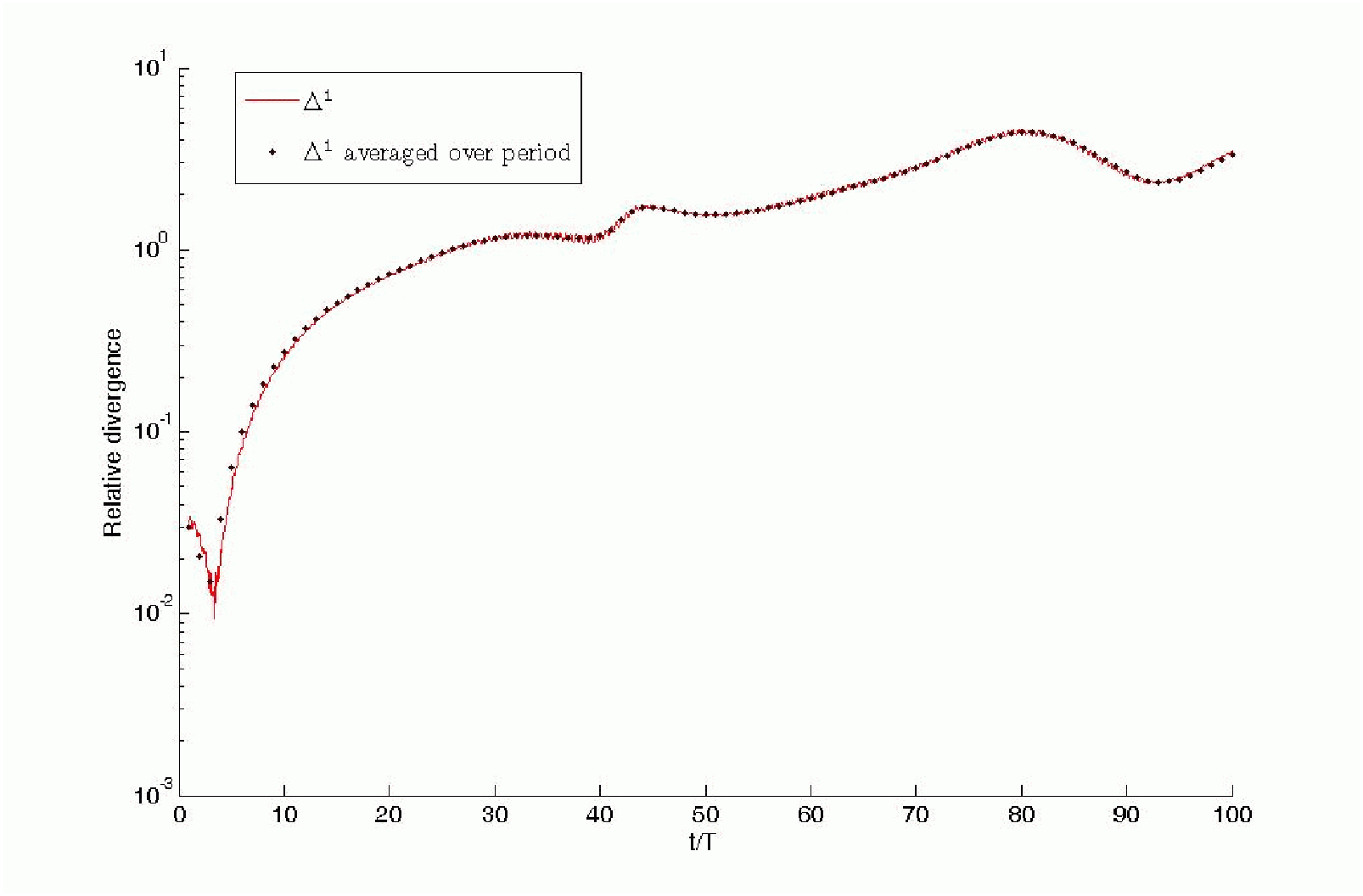}}
    \subfigure[$n=2, \hbar=0.032$]{
    \includegraphics[height=4cm,width=0.4\textwidth]{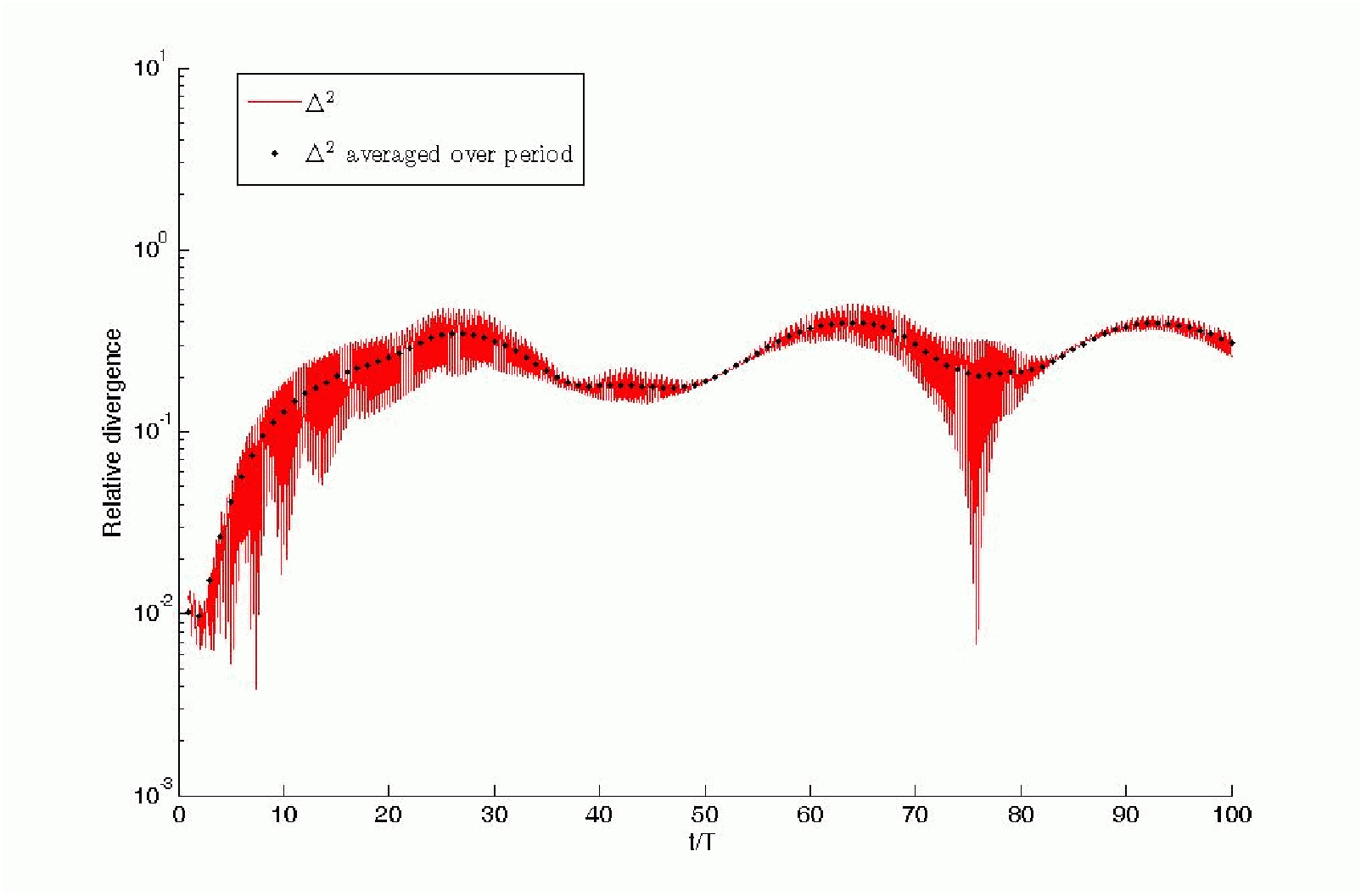}}
    \subfigure[$n=1, \hbar=0.008$]{
    \includegraphics[height=4cm,width=0.4\textwidth]{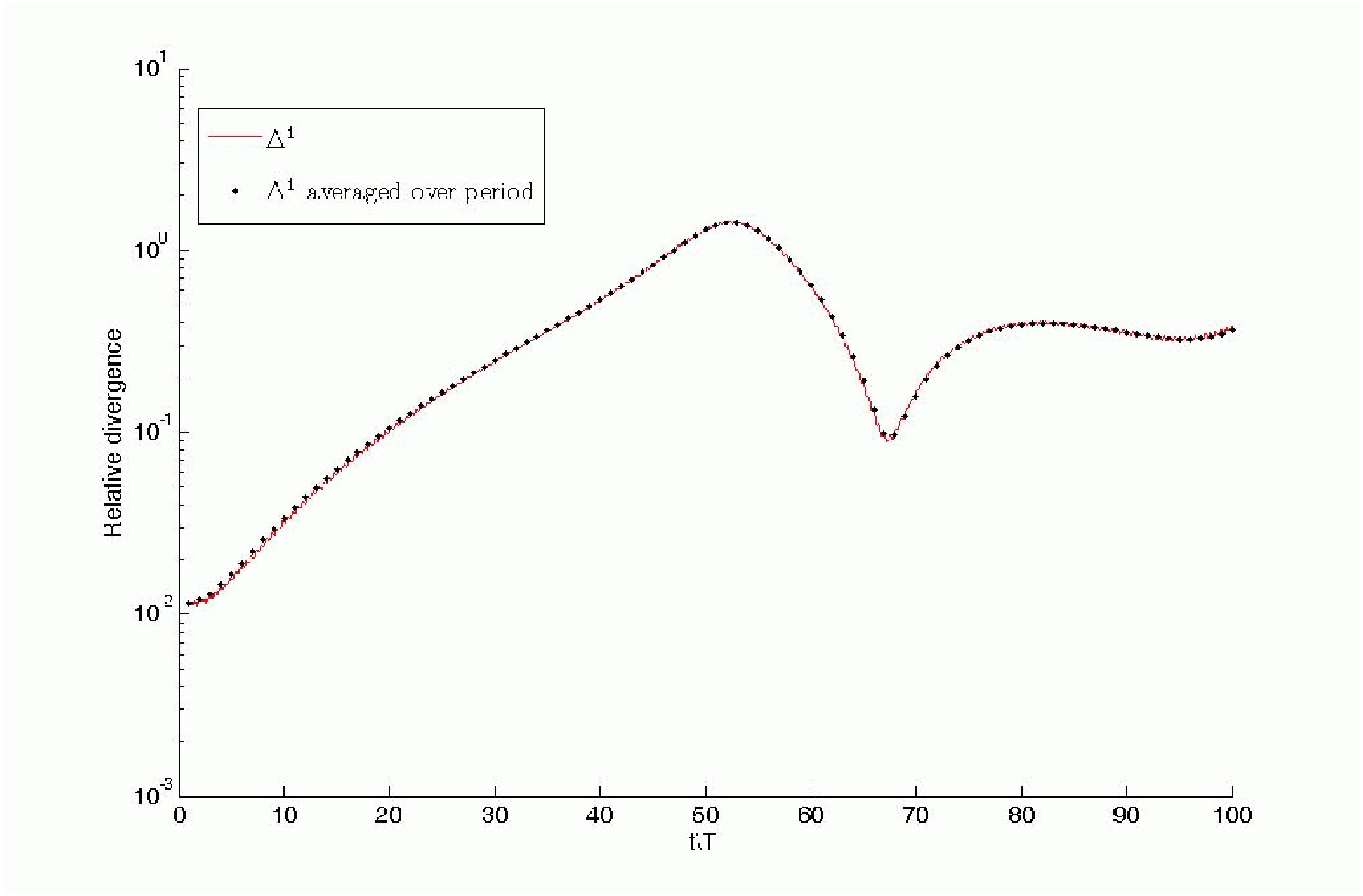}}
    \subfigure[$n=2, \hbar=0.008$]{
    \includegraphics[height=4cm,width=0.4\textwidth]{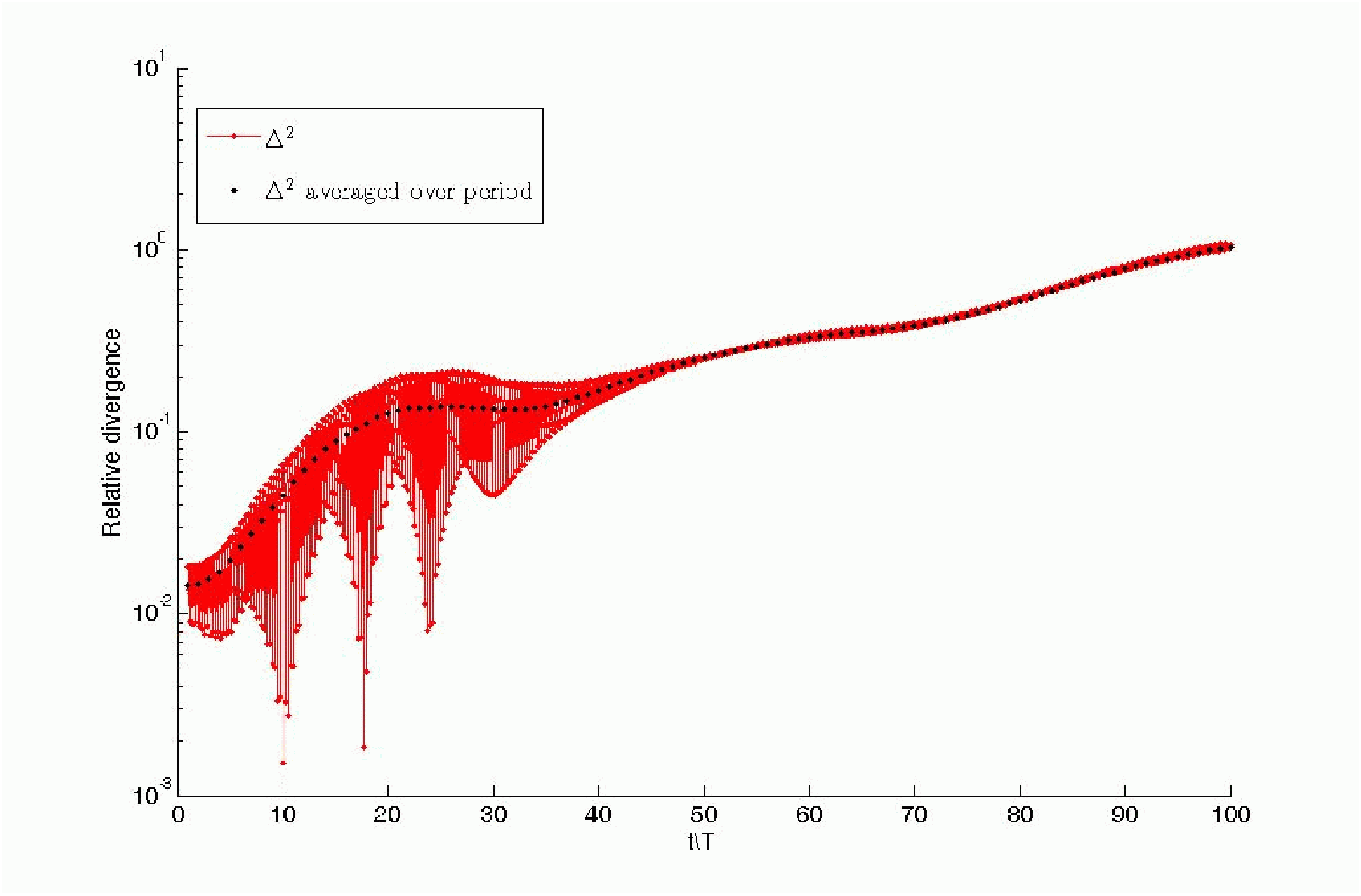}}
    \end{center}
    \caption{Relative divergence of quantum and classical expectation values
      for the isolated driven Duffing resonator, calculated using
      Eq.~(\ref{eq:divergence}). One cannot conclusively estimate the
      Ehrenfest times $t_E$ for the cases examined here. }
  \label{relative divergence hbar 0.2 and 0.8}
\end{figure}

We use the calculated phase-space densities in order to try to
estimate the time scale $t_E$ at which the quantum and classical
correspondence is broken. We do so by measuring the distance in phase
space between the quantum and classical expectation values of the
first and second moments of $x$ and $p$ as a function of time, scaled
by the distance of the classical phase-space point from the origin.
This measure, 
\begin{equation}\label{eq:divergence}
\Delta^{n} = \frac{\sqrt{(\langle
x^{n}\rangle-x_{cl}^{n})^{2}+(\langle
p^{n}\rangle-p_{cl}^{n})^{2}}}{\sqrt{(x_{cl}^{n})^{2}+(p_{cl}^{n})^{2}}},
\qquad n=1,2,
\end{equation}
gives us the relative divergence of the quantum expectation values
from the classical ones, where $\langle \cdot\rangle$ denotes the
quantum expectation value, and the subscript ${cl}$ denotes the
classical ensemble average. The results for $\Delta^1$ and $\Delta^2$
are shown in Fig.~\ref{relative divergence hbar 0.2 and 0.8} on a
semi-log scale as a function of time, along with their values averaged
over one period of the drive. For both values of the effective $\hbar$
that are shown it appears difficult to determine the Ehrenfest times
$t_E$. Even if one insists on making such a determination, it seems to
depend on the power of $x$ and $p$ being measured, and the resulting
estimates are inconclusive. Therefore, we cannot determine the
Eherenfest times in our model with any real certainty.

\section{Open System  -- Method of Calculation}
\label{open system dynamics}

We now couple the Duffing resonator to an external environment by
introducing a thermal bath. As usual, this coupling introduces
dissipation due to loss of energy from the resonator to the bath, as
well as fluctuations due to the random forces applied by the heat bath
to the resonator. We use different approaches to add this coupling to
either the classical or quantum-mechanical resonators, as outlined
below.

\subsection{Open classical dynamics -- Langevin equation} 
%and Fokker-Planck

In the classical calculation, dissipation is introduced using a
Langevin approach, adding two terms to the equation of motion for $p$.
The first is a velocity dependent friction force $-\gamma\dot{x}$, and
the second is a time dependent random force $\delta F(t)$ acting on
the resonator. Note that these additional terms are dimensionless and
measured using the units that were introduced earlier. In particular,
$\gamma$ is in fact the inverse of the dimensionless quality factor
$Q$. The random force is assumed to be a $\delta$-correlated Gaussian
white noise, satisfying
\begin{eqnarray}
\langle\delta F(t)\rangle = 0,\\
\langle\delta F(t)\delta F(t')\rangle = 2\gamma
k_{B}T_{env}\delta(t-t'),
\label{random force} 
\end{eqnarray}
where $k_{B}$ is Boltzman's constant, thus introducing
the bath temperature $T_{env}$.  The equations of motion become
\numparts
\begin{eqnarray}\label{classical eom open}
\dot{x} &=& p,\\
\dot{p} &=& -x-x^{3}-\gamma p+F\cos\omega t+\delta F(t),
\end{eqnarray}
\endnumparts and are integrated numerically, as in the case of the
isolated system.

\subsection{Open quantum dynamics -- Master equation}

To describe the evolution of a quantum Duffing resonator that is
coupled to an environment we can proceed either by using a quantum
Langevin approach~\cite{gardiner_zoller04}, or by coupling the
resonator to a bath of harmonic
resonators~\cite{gardiner_zoller04,louisell73}. The latter approach,
which we follow here, adds two terms to the Hamiltonian~(\ref{Quantum
  Hamiltonian})---a Hamiltonian for the bath and an interaction
Hamiltonian---resulting in
\begin{equation}\
H_{tot} = H_{sys}+H_{bath}+V.
\end{equation}

The von Neumann equation~(\ref{VN}) for the total density
operator $\rho_{tot}$, describing the resonator and the bath, is
\begin{equation}\label{von neumann}
\dot{\rho}_{tot} = \frac{1}{i\hbar}[H_{tot},\rho_{tot}],
\end{equation}
where the resonator alone is now described by the reduced density
operator, obtained by tracing out the bath degrees of freedom,
\begin{equation}
\rho_{sys}(t)=Tr_{bath}\left\{\rho_{tot}(t)\right\}.
\end{equation}
We use a standard interaction Hamiltonian~\cite{santamore04} of the
Caldeira-Leggett~\cite{caldeira_leggett83} type in the rotating-wave
approximation,
\begin{equation}\label{cald_legget_interaction}
V=\sum_{i}(\kappa_{i}b_{i}a^{\dagger}+\kappa_{i}^{*}b_{i}^{\dagger}a),
\end{equation}
where the $\kappa_{i}$ are bilinear coupling constants, the $b_{i}$
are annihilation operators acting on the bath oscillators, and $a$ is
the annihilation operator of the Duffing resonator~(\ref{eq:ladder}
and $b$). Note that a different choice of coupling other than
Eq.~(\ref{cald_legget_interaction}) may lead to a different Master
equation. 

By assuming the interaction to be weak, and by employing the Markov
approximation which assumes that the bath has no memory, we obtain a standard
master equation~\cite{santamore04,gardiner_zoller04,louisell73}, 
\begin{eqnarray}\label{master equation}\nonumber
  \dot\rho_{sys} &= &\frac{1}{i\hbar}[H_{sys},\rho_{sys}] -
  \frac{\gamma}{2}(1+\bar{n})(a^{\dagger}a\rho_{sys} +
  \rho_{sys}a^{\dagger}a - 2a\rho_{sys}a^{\dagger})\\ 
  &- &\frac{\gamma}{2}\bar{n}(aa^{\dagger}\rho_{sys} +
  \rho_{sys}aa^{\dagger} - 2a^{\dagger}\rho_{sys}a),
\end{eqnarray}
where $\bar{n}\equiv(e^{\hbar\Omega/k_{B}T_{env}}-1)^{-1}$ is the
Bose-Einstein distribution through which $T_{env}$ is introduced,
$\gamma\equiv2\pi g(\Omega)|\kappa(\Omega)|^{2}$ is interpreted as the
damping rate, $g(\Omega)$ is the density of states of the bath,
evaluated at the resonator's natural frequency, and $\kappa(\Omega)$
is the coupling constant at the resonator's natural frequency. We
evaluate these two variables at the natural frequency $\Omega=1$,
because the shift in the resonator's frequency due to nonlinearity is
assumed small.

The resulting master equation~(\ref{master equation}) is of a Lindblad
form~\cite{gardiner_zoller04,schlosshauer07,spohn80}, which in the
Schr\"{o}dinger picture is given by
\begin{equation}\label{lindblad}
\dot\rho = \frac{1}{i\hbar}[\tilde{H},\rho] -
\frac{1}{2}\sum_{m}(C_{m}^{\dagger}C_{m}\rho 
+ \rho C_{m}^{\dagger}C_{m}) + \sum_{m}C_{m}\rho C_{m}^{\dagger},
\end{equation}
where $\tilde{H}$ may differ from $H_{sys}$, for example due to a Lamb
shift of the resonator's natural frequency. The non-unique operators
$C_{m}$ and $C_{m}^{\dagger}$ operate on the Hilbert space of the
resonator. Identifying the Lindblad form enables us to avoid solving
the master equation directly---a solution which demands extensive
computation. Instead, we use the \textit{Monte-Carlo wave function
  method} (MCWF) \cite{MCWF2,MCWF}, which is computationally more
efficient. It is based on computing many time evolutions of the
initial quantum state, so-called quantum trajectories, with a
non-hermitian Hamiltonian that implements the stochastic action of the
bath and is derived from $H_{sys}$. Subsequent averaging over the
different trajectories yields the correct time evolution of the
density operator that satisfies the master equation~(\ref{master
  equation}). We carry out the MCWF numerically using MATLAB.

\section{Open System at $T_{env}=0$ -- Results}
\label{open system T zero results}

\begin{figure}[tbh]
  \begin{center}
    \subfigure[Initial coherent state, $t=0$]{
      \includegraphics[clip=true, viewport=0.6cm 2.4cm 13.6cm
      8.3cm,width=0.45\textwidth]{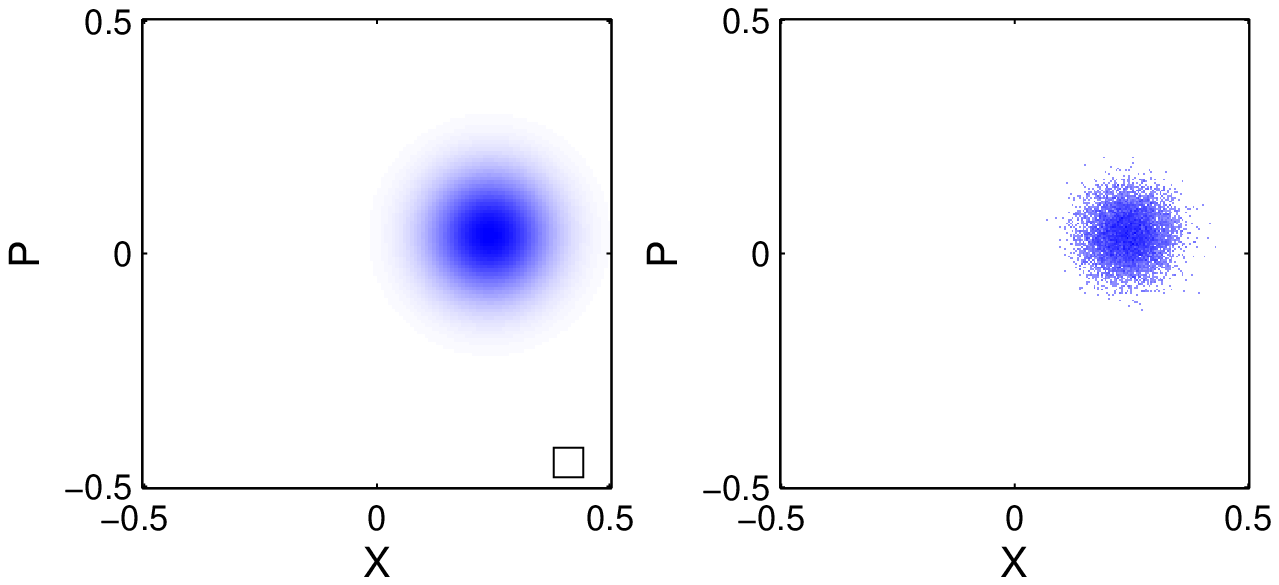}}
    \subfigure[$t=10T$]{
      \includegraphics[clip=true, viewport=0.6cm 2.4cm 13.6cm
      8.3cm,width=0.45\textwidth]{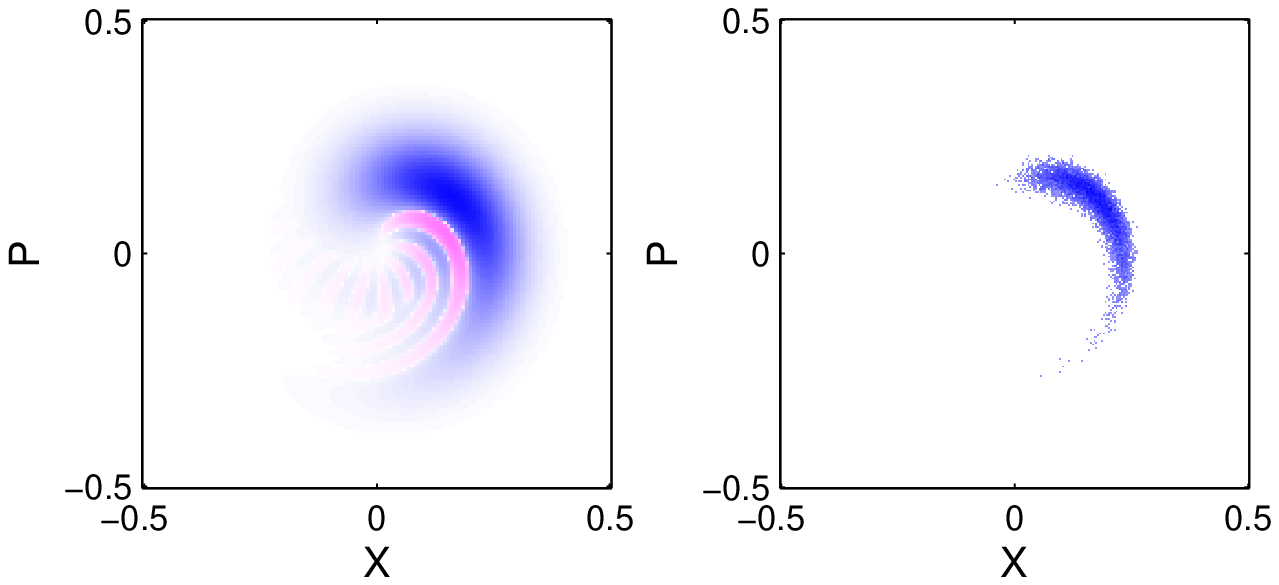}}
    \subfigure[$t=80T$]{
      \includegraphics[clip=true, viewport=0.6cm 2.4cm 13.6cm
      8.3cm,width=0.45\textwidth]{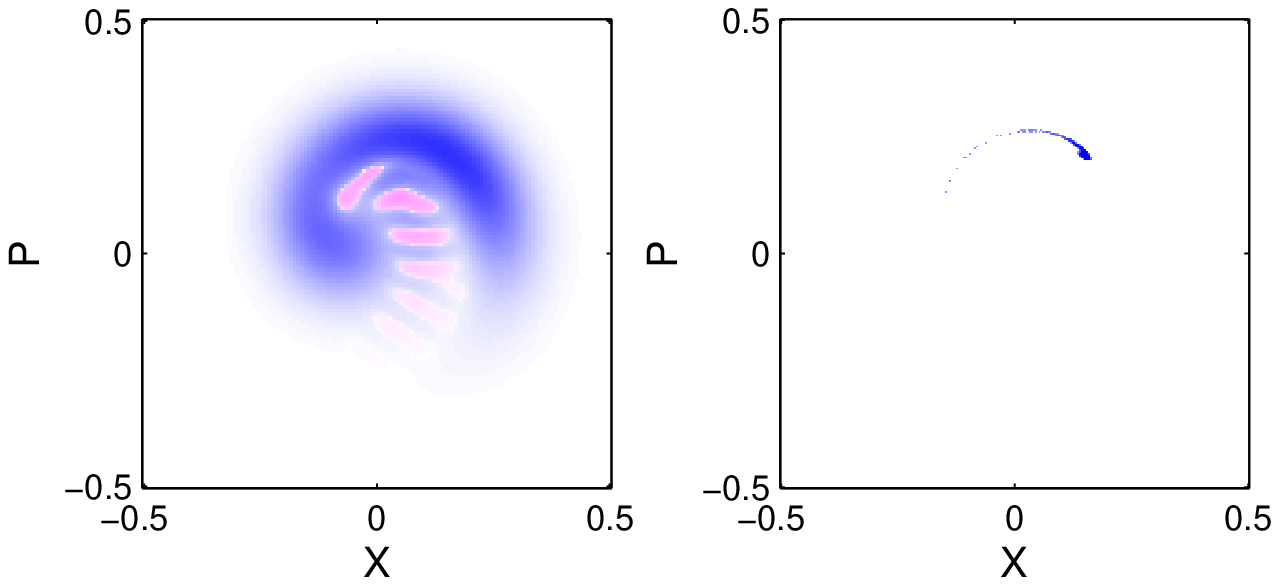}}
    \subfigure[$t=300T$]{
      \includegraphics[clip=true, viewport=0.6cm 2.4cm 13.6cm
      8.3cm,width=0.45\textwidth]{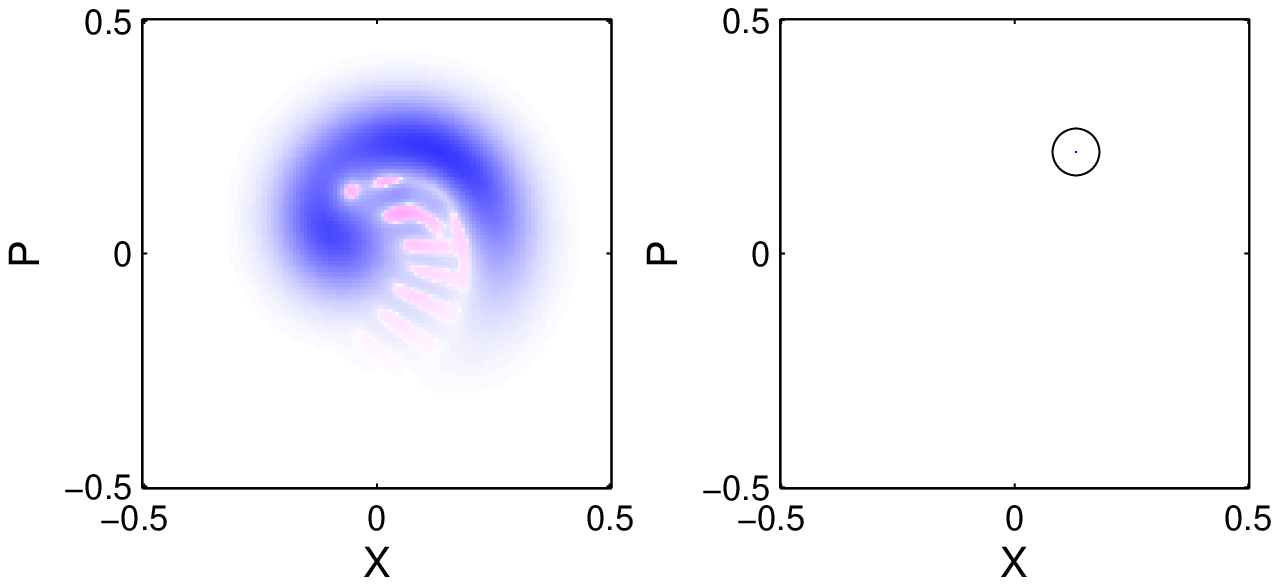}}
  \end{center}
  \caption{A driven Duffing resonator coupled to a zero-temperature heat bath
    with scaled $\hbar=0.004$ (other parameters are listed in the
    text). Wigner functions (left) and classical phase-space
    distributions (right) of the initial minimal Gaussian wave
    packet---located within the basin of attraction of the
    large-amplitude dynamical state---and its evolution at three later
    times.  The large-amplitude fixed point, towards which the
    classical distribution flows, is encircled in Figure (d).  A
    square of area $\hbar$ is shown at the bottom right corner of the
    initial state plot.  The functions are scaled by $f\rightarrow
    f^{1/4}$ for a better color contrast.  }
  \label{nonlinear open kT0 hb0.1}
\end{figure}

We begin by considering the coupling of the driven Duffing resonator
to a zero-temperature bath. Again, we choose the values of the
different parameters $F=0.015$, $\omega=1.018$, and $\gamma=0.01$,
such that the driven nonlinear resonator is operating in its
bistability regime.  For $T_{env}=0$, in the quantum dynamics the
average phonon number $\bar{n}=0$, and in the classical dynamics we
expect to see damping without the smearing effect of a random force.
Thus, the ability of the environment to induce transitions between the
dynamic states of the resonator is suppressed. We look
at a value of $\hbar=0.004$, as shown in Fig.~\ref{nonlinear open kT0 hb0.1},
where the initial state is placed within the basin of attraction of
the large amplitude classical stable solution. The Gaussian classical 
phase-space distribution all flows, as expected, to a single
point---the fixed point corresponding to the large-amplitude dynamical
state. 

In the quantum dynamics, at short times we again see a general
positive outline of the Wigner function which is similar to the
classical density, as for the isolated Duffing resonator in
Section~\ref{closed system results}, but it quickly deviates from the
classical distribution. The uncertainly principle prevents it from
shrinking to a point as in the classical dynamics. More importantly,
we clearly observe that the Wigner distribution has substantial weight
around the state of small-amplitude oscillations, which is
inaccessible classically for the chosen initial conditions at
$T_{env}=0$. \emph{The quantum resonator can still switch between the two
stable dynamical states, even though the classical resonator cannot.}
This switching takes place either via tunneling---in analogy with the
macroscopic quantum tunneling between two equilibrium states of a
static system~\cite{Carr01}---or via quantum
activation~\cite{dykman06,dykman07}, although at this point it is
impossible for us to distinguish between these two processes.

\goodbreak
\section{Open System at $T_{env}>0$ -- Results}
\label{open system T nonzero results}

\begin{figure}[tbh]
  \begin{center}
    \subfigure[Initial coherent state, $t=0$]{
      \includegraphics[clip=true, viewport=0.6cm 2.4cm 13.6cm
      8.3cm,width=0.45\textwidth]{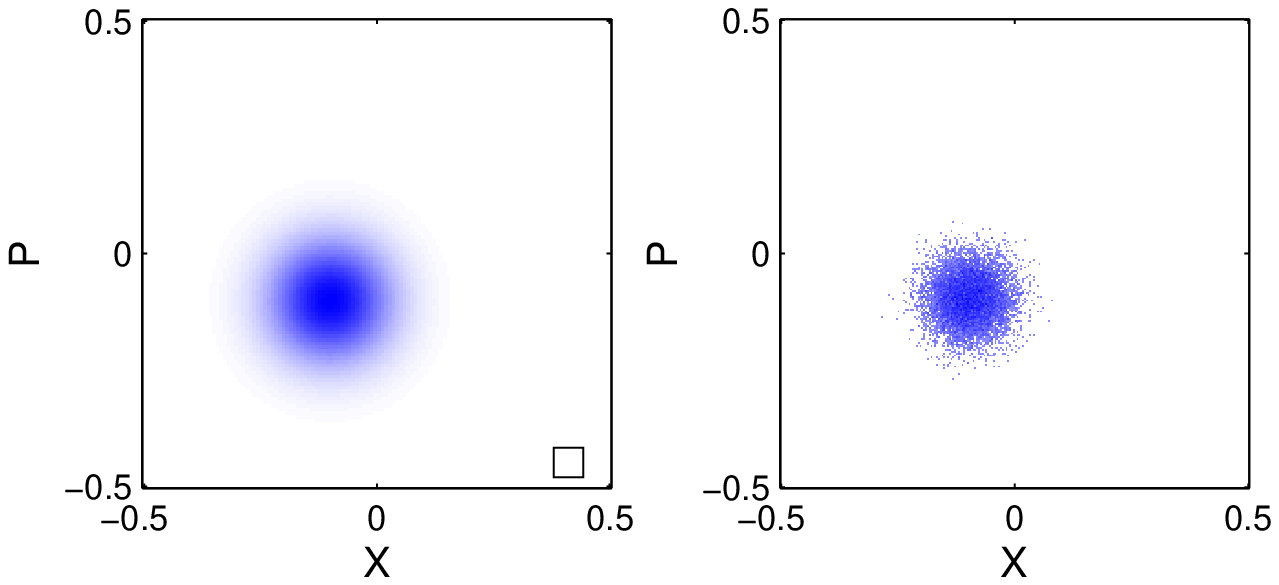}}
    \subfigure[$t=30T$]{
      \includegraphics[clip=true, viewport=0.6cm 2.4cm 13.6cm
      8.3cm,width=0.45\textwidth]{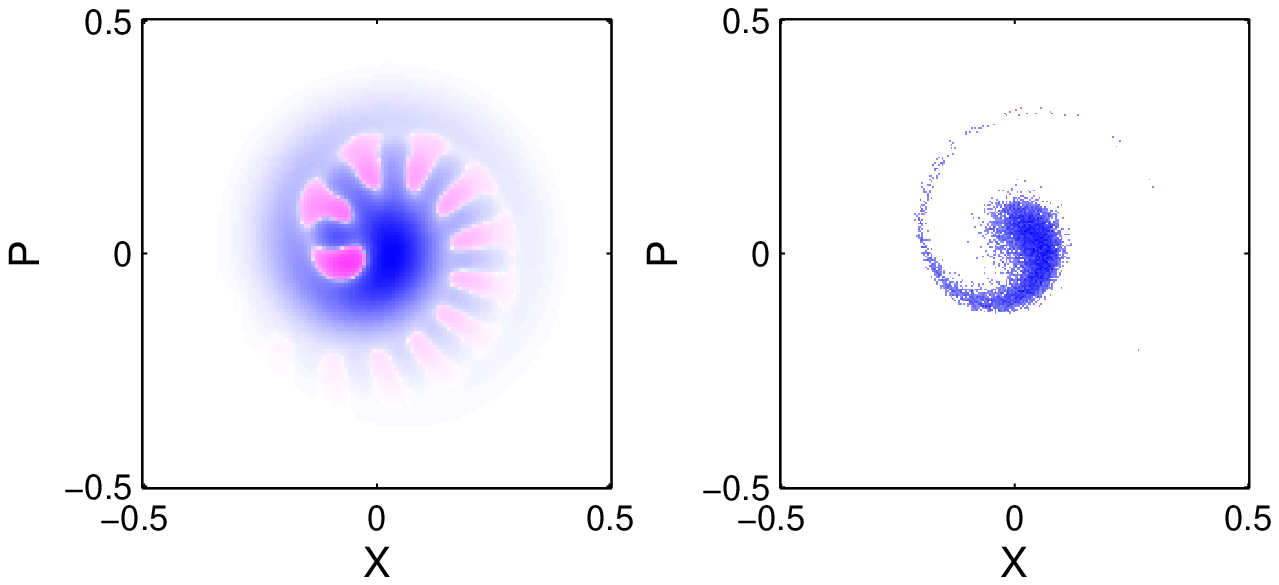}}
    \subfigure[$t=180T$]{
      \includegraphics[clip=true, viewport=0.6cm 2.4cm 13.6cm
      8.3cm,width=0.45\textwidth]{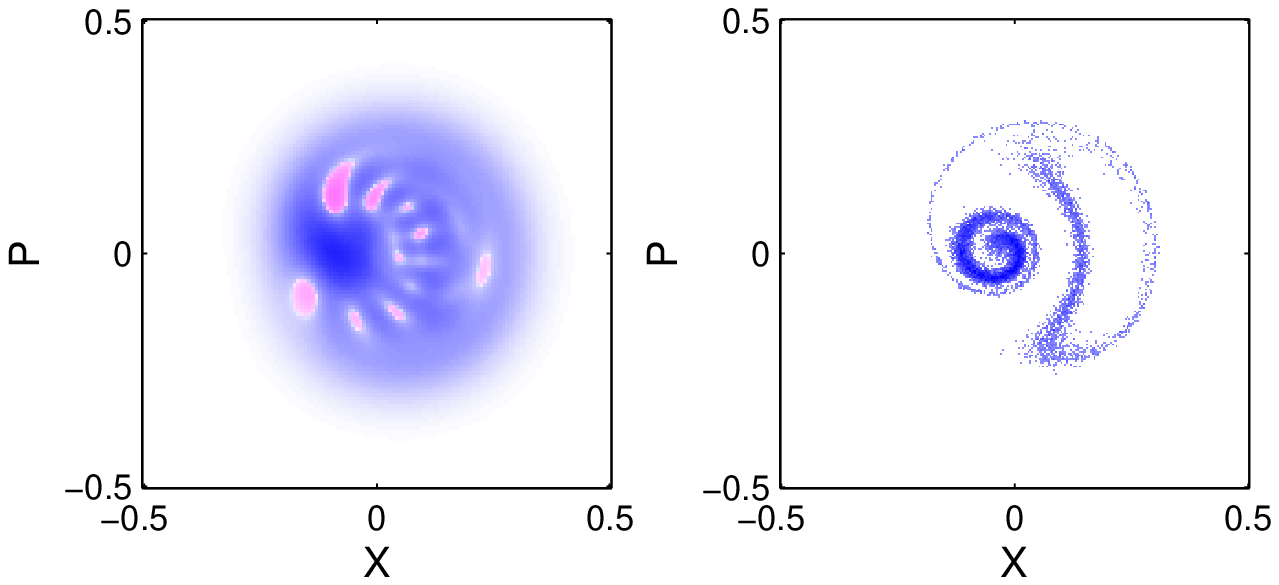}}
    \subfigure[$t=800T$]{
      \includegraphics[clip=true, viewport=0.6cm 2.4cm 13.6cm
      8.3cm,width=0.45\textwidth]{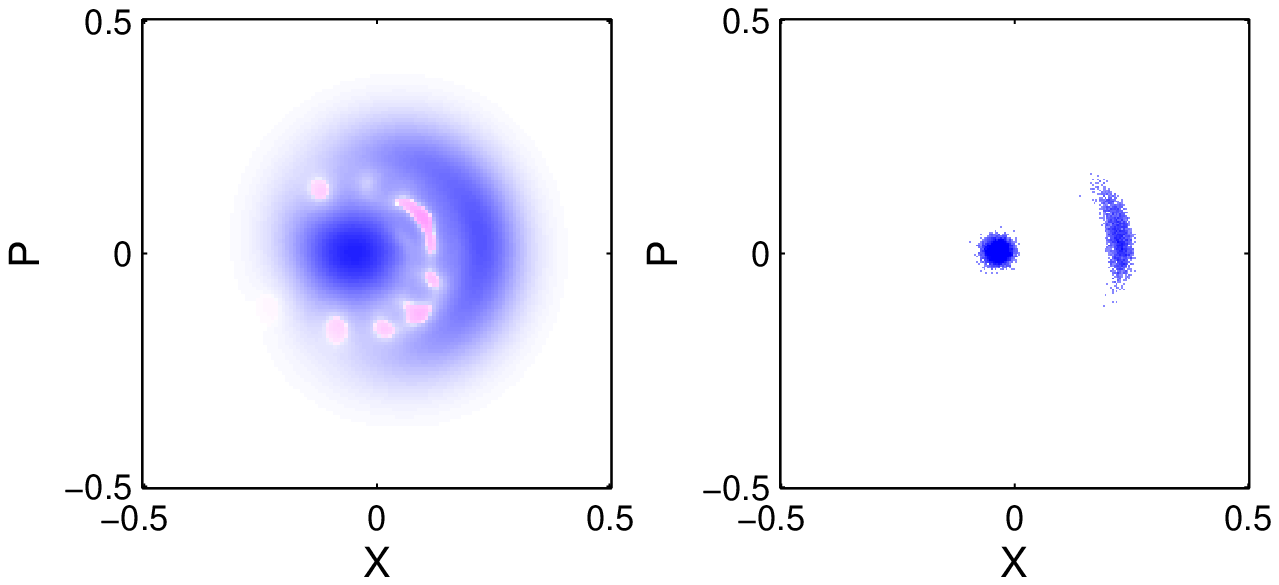}}
  \end{center}
  \caption{The driven Duffing resonator coupled to a heat bath at
    $k_{B}T_{env}=2\hbar\Omega$ with $\hbar=0.004$. Wigner function
    (left) and classical distribution (right) of an initial minimal
    Gaussian wave packet---straddling the separatrix between the
    regions in phase space flowing to the two stable states---and its
    evolution at three later times.  A square of area $\hbar$ is shown
    at the bottom right corner of the initial state plot.  The
    functions are scaled by $f\rightarrow f^{1/4}$ for a better color
    contrast.  }
  \label{nonlinear open kT0.2 hb0.1}
\end{figure}

\begin{figure}[hbt]
  \begin{center}
    \subfigure[Initial coherent state, $t=0$]{
      \includegraphics[clip=true,viewport=0.6cm 2.3cm 14cm
      8.5cm,width=0.45\textwidth,height=3cm]{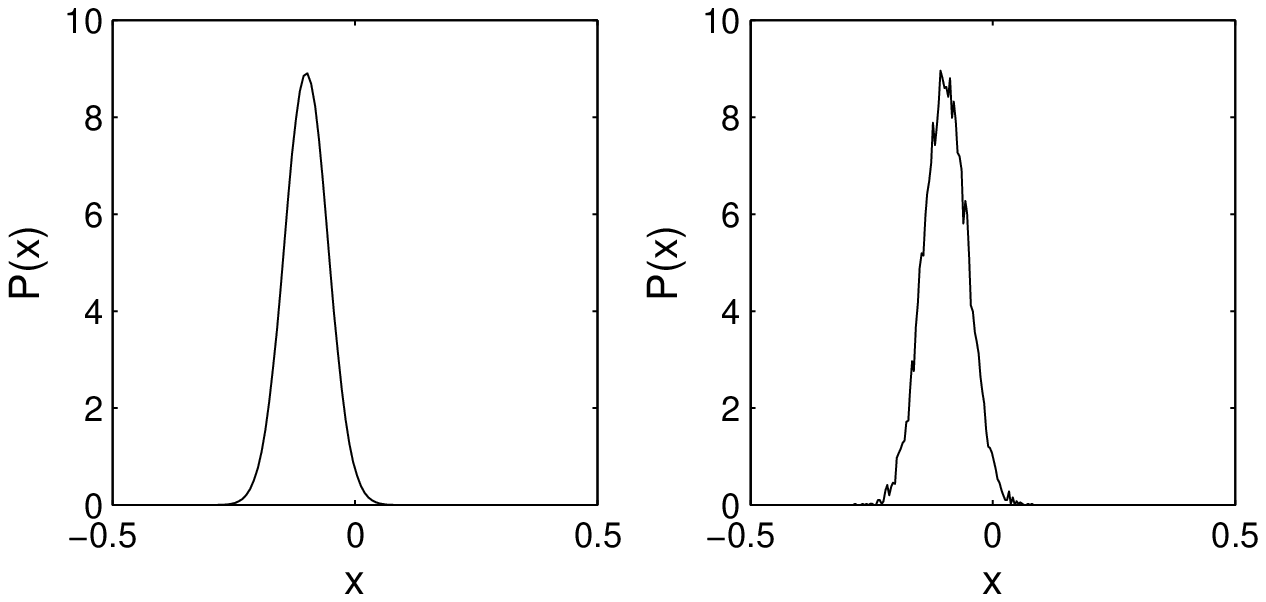}}
    \subfigure[$t=30T$]{
      \includegraphics[clip=true,viewport=0.6cm 2.15cm 14cm
      8.5cm,width=0.45\textwidth,height=3cm]{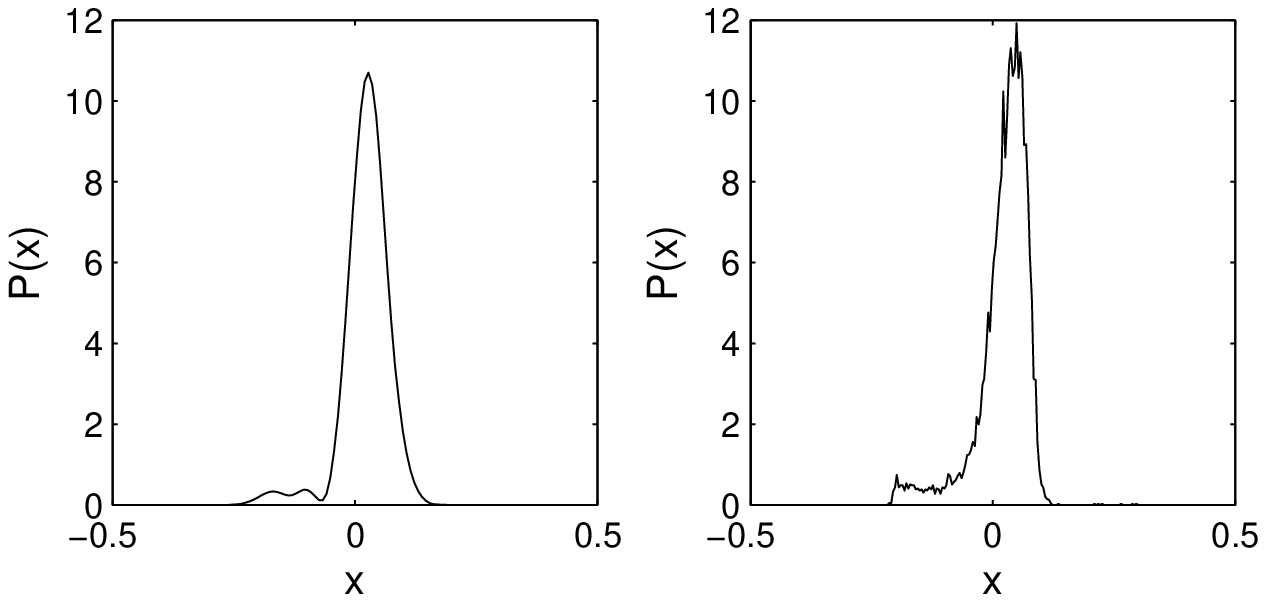}
      \vspace{5.0cm}} 
    \subfigure[$t=180T$]{
      \includegraphics[clip=true,viewport=0.6cm 2.3cm 14cm
      8.5cm,width=0.45\textwidth,height=3cm]{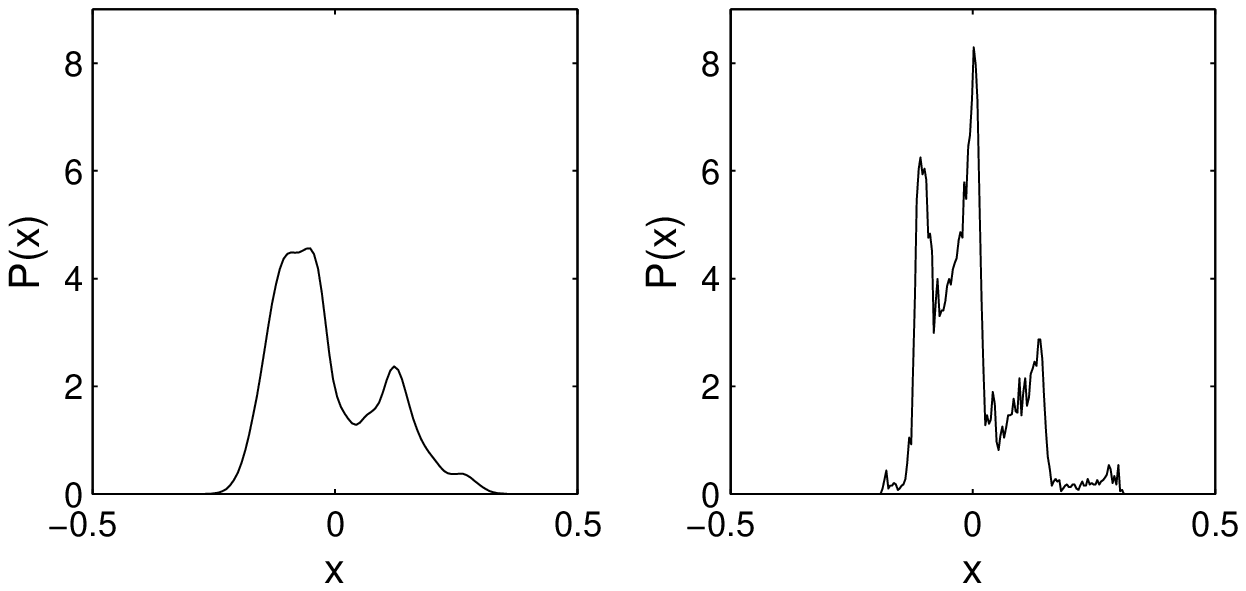}}
    \subfigure[$t=800T$]{
      \includegraphics[clip=true,viewport=0.5cm 2.15cm 14cm
      8.5cm,width=0.45\textwidth,height=3cm]{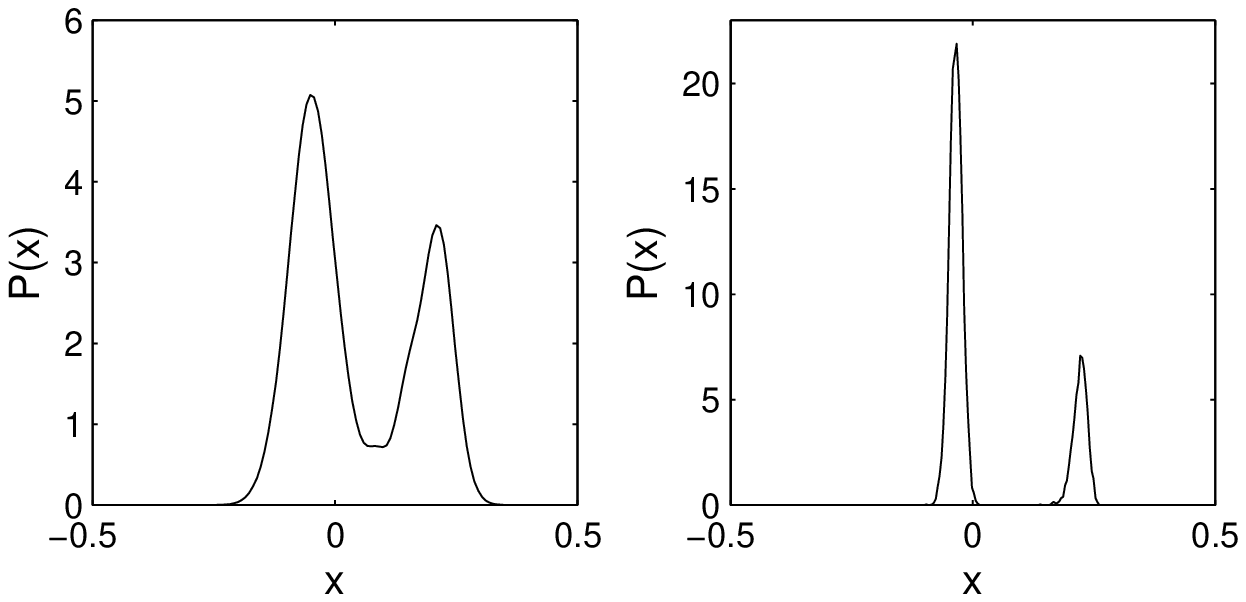}}
  \end{center}
  \caption{Probability densities $P(x)$ for the same four states shown in
    Fig.~\ref{nonlinear open kT0.2 hb0.1}, obtained by integrating the
    phase-space distributions in the $p$ direction. The $y$ axis is
    scaled differently for better visualization.}
  \label{nonlinear open kT0.2 hb0.1PD}
\end{figure}

Figure~\ref{nonlinear open kT0.2 hb0.1} shows the calculated results
for a Duffing resonator with $\hbar=0.004$, coupled to a heat bath at
a finite temperature $k_{B}T_{env}=2\hbar\Omega$. This temperature is
obtained by adding a fluctuating force according to the
fluctuation-dissipation relation~(\ref{random force}). To avoid having
a random force whose magnitude dominates the dynamics we reduce the
damping rate to $\gamma=0.001$, thus requiring a longer time for the
resonator to reach its final steady state. The remaining parameters
are chosen to be $F=0.006$ and $\omega=1.016$ to ensure that the
resonator is still operating in the bistability regime.

Here we choose the initial Gaussian state to straddle the separatrix
between regions in phase space that flow to the two stable states. The
interference pattern that develops for short times within the general
positive outline in the Wigner function is soon destroyed by
decoherence, and becomes jittery in space and time. At long times,
both distributions peak around the two stable states, nevertheless
they differ significantly. The classical density is tightly localized
around the two solutions with no overlap, indicating that
\emph{$T_{env}$ is too small to induce thermal switching of the
  classical resonator between the two states,} as was observed in a
recent experiment~\cite{aldridge05}. The Wigner function, on the other
hand, is spread out in phase space, indicating that \emph{$\hbar$ is
  sufficiently large for the quantum resonator to switch between the
  two states} via tunneling or quantum
activation~\cite{dykman06,dykman07}.  This is demonstrated more
clearly in Fig.~\ref{nonlinear open kT0.2 hb0.1PD}, which shows the
probability distributions $P(x)$, that are obtained by integrating the
phase-space distributions in Fig.~\ref{nonlinear open kT0.2 hb0.1} in
the $p$ direction.

\begin{figure}[bt]
    \begin{center}
    \subfigure[\ Phase space densities, $t=800T$]{
      \includegraphics[clip=true,viewport=0.6cm 2.3cm 14cm
      8.5cm,width=0.4\textwidth,height=3cm]{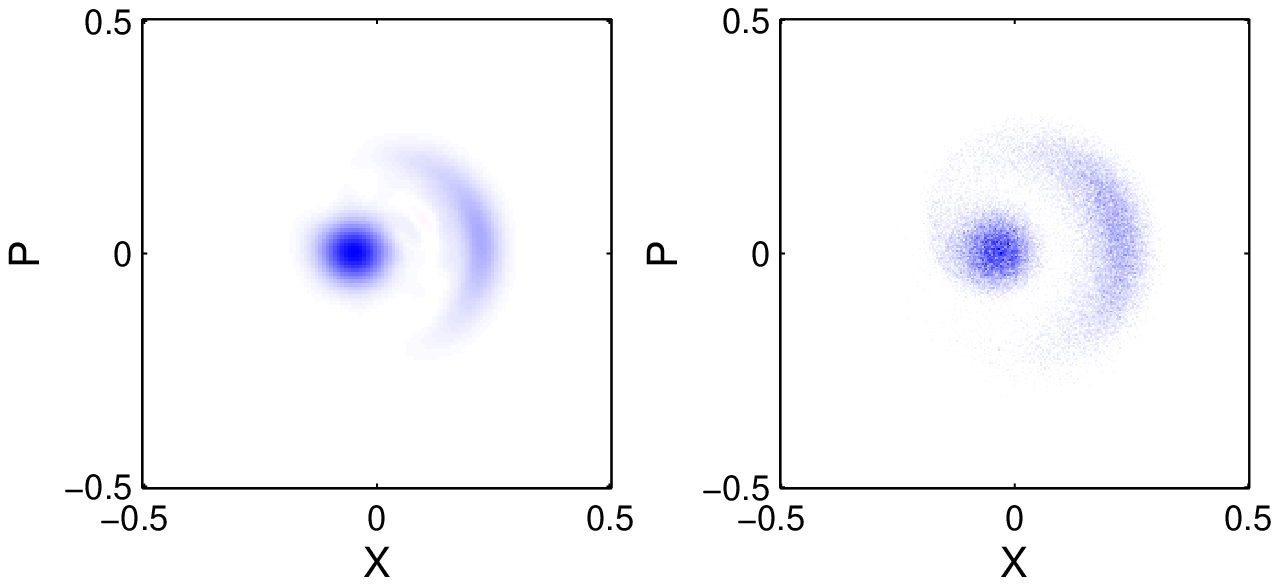}}
    \subfigure[\ Probability densities of $x$, $t=800T$]{
      \includegraphics[clip=true,viewport=0.6cm 2.3cm 14cm
      8.5cm,width=0.4\textwidth,height=3cm]{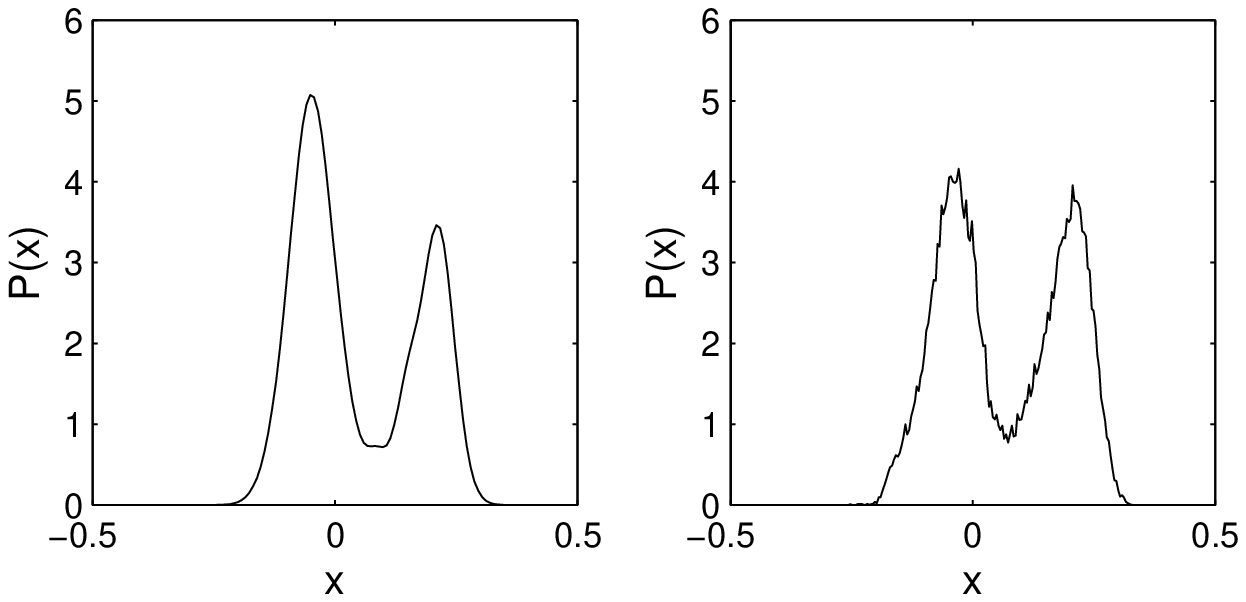}}
    \end{center}
    \caption{As in Figs.~\ref{nonlinear open kT0.2 hb0.1}(d)
      and~~\ref{nonlinear open kT0.2 hb0.1PD}(d), only that the
      classical calculation (right) is for
      $k_{B}T_{env}=17\hbar\Omega$, yielding (a) similar phase-space
      distributions; and (b) similar probability densities. Both
      distributions in (a) are left unscaled.}
    \label{effective temperature}
\end{figure}

Further analysis shows that only for temperatures as high as
$k_{B}T=17\hbar\Omega$ does the classical phase-space distribution
become as wide as the Wigner function is at $k_{B}T=2\hbar\Omega$, as
demonstrated in Fig.~\ref{effective temperature}.  Thus, in a real
experiment, evidence for quantum-mechanical dynamics can be
demonstrated as long as temperature and other sources of noise can be
controlled to better than an order of magnitude. Note also that minor
differences exist between the high-temperature classical distribution
and the low-temperature quantum distribution (\textit{e.g.} the
relative heights of the two peaks in the probability density
$P(x)$, as seen in Fig.~\ref{effective temperature}(b)), which may
assist in distinguishing between the two possibilities.

%% One phenomenon which exists only in the quantum dynamics is the
%% nonzero probability of finding the resonator in a position in
%% between the two classically stable solutions, where classically this
%% probability is zero. Another such phenomenon is the negative part of
%% the Wigner function, which is not negligible though it does not
%% incorporate a standing-wave-like interference pattern and is about
%% an order of magnitude smaller than the peak of the Wigner
%% function.

As we lower the temperature, for example, down to
$k_{B}T_{env}=\hbar\Omega$, we see further deviations of the quantum
dynamics from the classical one.  Figure~\ref{nonlinear open final
  wigner 3D} shows the unscaled Wigner functions of the state at
$t=800T$ for the two temperatures. In both cases the function is
almost all positive with small negative parts, and are peaked around
the two classical stable solutions. Nevertheless, one sees that the
similarity to the classical distribution is evidently higher for the
higher temperature.

\begin{figure}[hbt]
    \begin{center}
    \subfigure[$k_{B}T_{env}=\hbar\Omega$]{
      \includegraphics[clip=true,viewport=0.6cm 2.1cm 15.5cm
      12.5cm,width=0.45\textwidth]{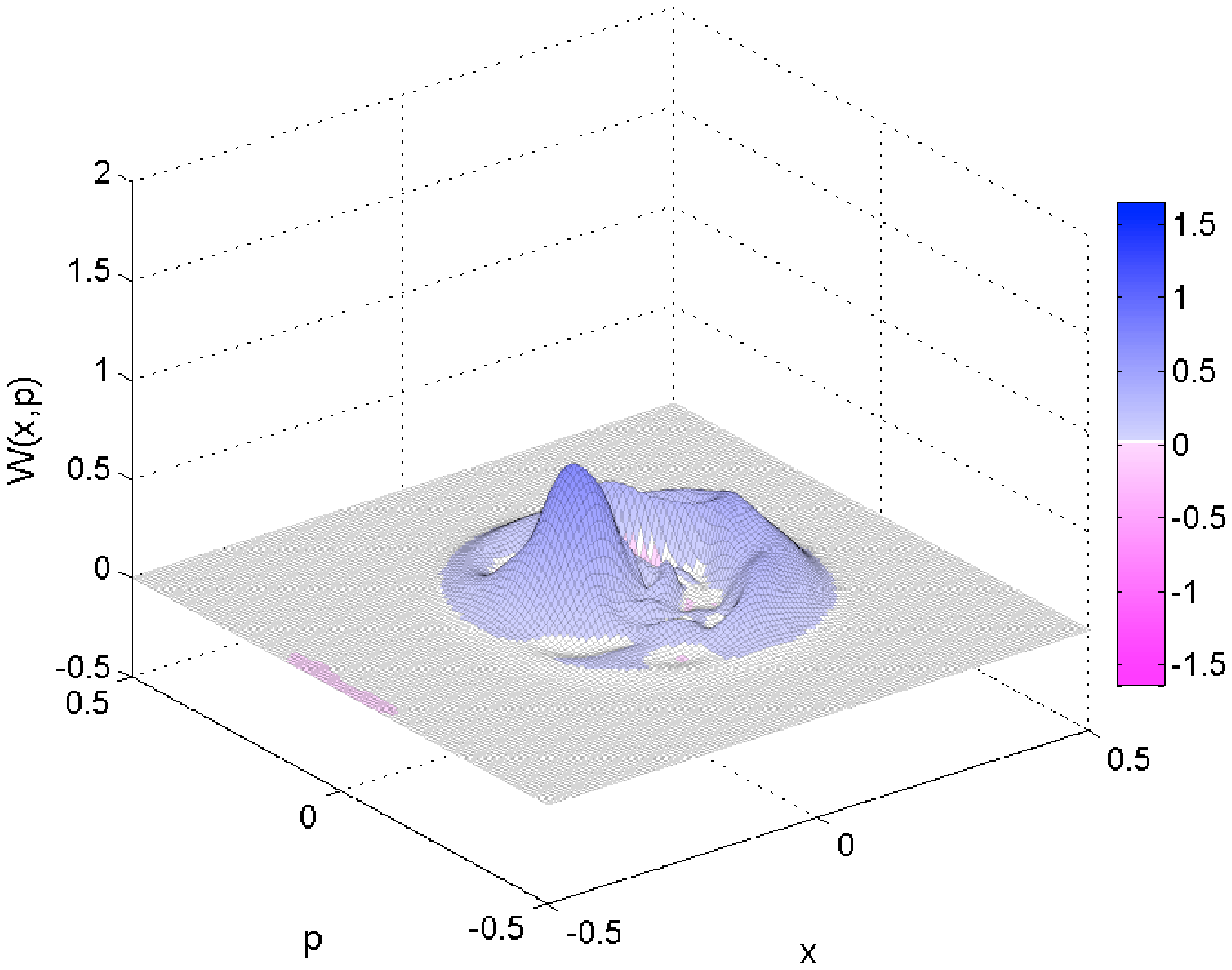}}
    \subfigure[$k_{B}T_{env}=2\hbar\Omega$]{
      \includegraphics[clip=true,viewport=0.6cm 2.1cm 15.5cm
      12.5cm,width=0.45\textwidth]{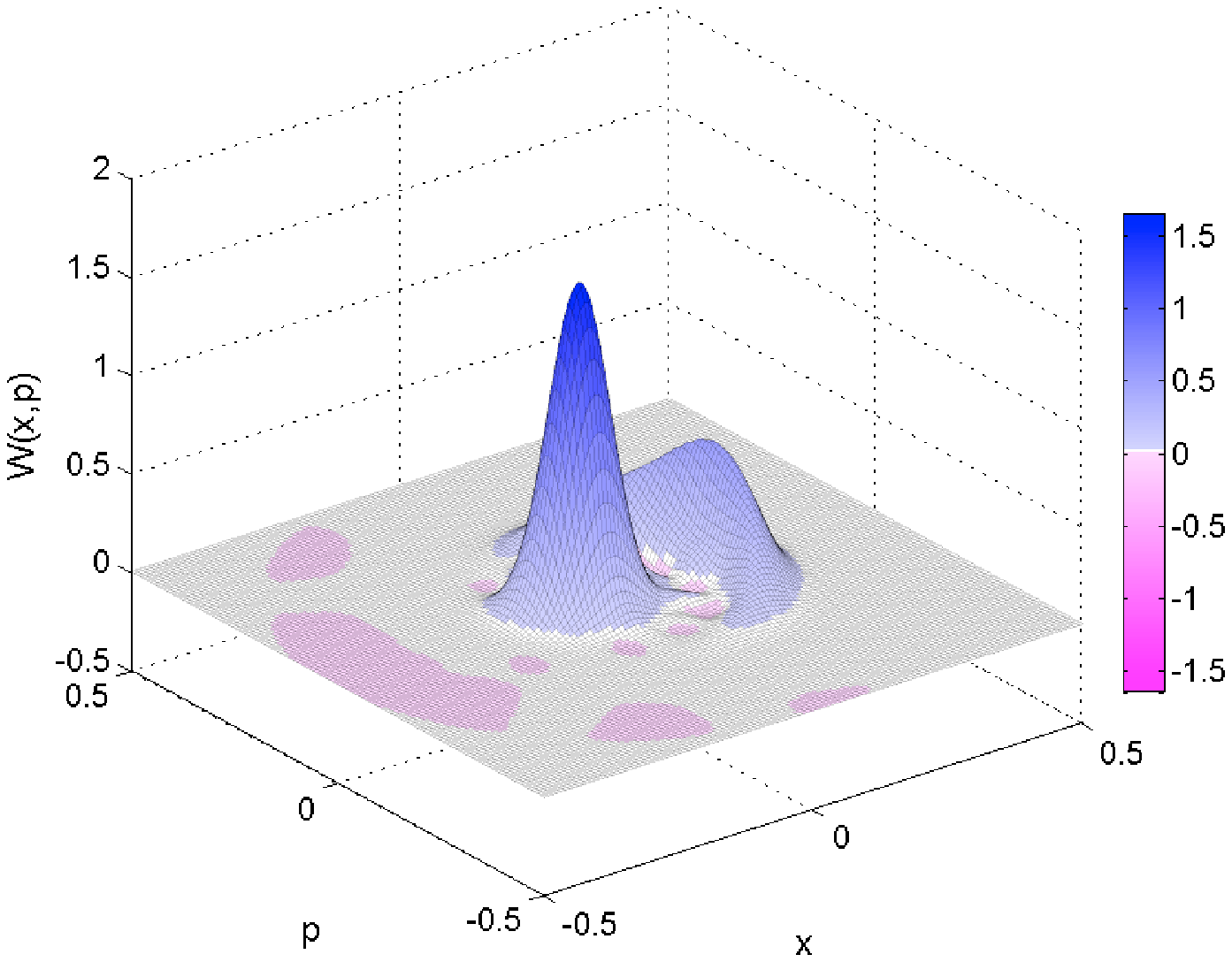}}
    \end{center}
    \caption{The driven Duffing resonator coupled to a heat bath at
      $k_{B}T_{env}=\hbar\Omega$ and at $k_{B}T_{env}=2\hbar\Omega$
      with $\hbar=0.004$. Unscaled Wigner functions of the state at
      $t=800T$ are plotted for the two temperatures. As in
      Fig.~\ref{nonlinear open kT0.2 hb0.1}, the initial Gaussian state
      straddles the separatrix so that the classical distribution splits
      towards both of the two stable dynamical states. With the
      decrease in temperature one sees further deviation away from
      classical behavior.}
    \label{nonlinear open final wigner 3D}
\end{figure}

\goodbreak
\section{Discussion}
\label{conclusion}

We have calculated the classical evolution and the quantum evolution
of a driven Duffing resonator, whether isolated or coupled to an
environment. For the case of an isolated resonator---which is less
relevant for experiment, yet still interesting theoretically---we have
confirmed that the quantum and classical evolutions are essentially
different, and that the transition from one picture to the other does
not take place in a simple manner. The non-analytic nature of the
limit $\hbar\to0$ appears in the form of strong quantum interference
patterns in the Wigner function, describing the quantum evolution,
which are never suppressed in the absence of coupling to an
environment. In addition, one observes an infinitely fine structure in
the classical phase-space distribution, which is absent from the
quantum distribution, owing to the uncertainty principle that smears
any structure on the scale of $\hbar$. We have shown that these
differences can be resolved by introducing a constraint on the
experimental resolution. An uncertainty in the measurement on the order
of $\hbar$ can both smear the fine structure in the classical
distribution and average out the interference pattern in the quantum
Wigner function, giving rise to similarly-looking phase-space
distributions, as demonstrated in Fig.~\ref{averaged wigner}.

In all cases investigated here---whether for an isolated resonator or
one that is coupled to an environment---we have found that at very
early stages of the evolution the quantum Wigner function and the
classical phase-space distribution agree with each other. As can be
seen in Figs.~\ref{nonlinear close 0.2}(b), \ref{nonlinear open kT0
  hb0.1}(b), and \ref{nonlinear open kT0.2 hb0.1}(b), at early times
the Wigner function contains a positive backbone which closely
resembles the classical distribution, even though a quantum
interference pattern develops within this backbone. One expects that
up to some Ehrenfest time there will be a corresponding agreement
between quantum observables and classical averages. Nevertheless, even
though such agreement is observed we could not consistently and
conclusively estimate the Ehrenfest time.

Only after coupling the driven Duffing resonator to an environment at
temperature $T_{env}$ was it possible to search for the regime of
interest that we set out to find at the beginning of this work. We
wanted to find a regime that would enable us to observe the first
deviations away from classical behavior as we pass through the
so-called classical-to-quantum transition. We have found this regime
by gradually decreasing the physical dimensions of the system, or
equivalently by increasing the effective value of $\hbar$, as measured
by the physical scale $m\Omega d^2$, set by the parameters of the
system. We have found an appropriate set of parameters---$\hbar\approx
10^{-3}$, $k_BT_{env}=2\hbar\Omega$, and $Q\approx 10^3$ with a
corresponding forcing amplitude that puts the Duffing resonator in its
bistable regime---for which the quantum dynamics, as it appears in the
Wigner function, looks very much like classical dynamics, but at the
same time shows clear signatures that it is not classical.

We should ask how far we are from reaching such parameters in
realistic systems with current technology. Reaching quality factors on
the order of hundreds to thousands is quite routine with standard
suspended elastic beam resonators and somewhat optimistic with
suspended nanotubes, while exciting these resonators into their
nonlinear regime and observing hysteresis due to bistability is very
common. Reaching the required effective $\hbar\approx10^{-3}$ implies
that one needs $m\Omega d^2\approx10^{-31}$ m$^2$kg/s. For a NEMS
beam~\cite{ina05,ina06} with $d\approx 10$nm and $m\approx10^{-18}$kg,
vibrating at $\Omega\approx10^8$Hz, we obtain $m\Omega
d^2\approx10^{-26}$ m$^2$kg/s, which is 5 orders of magnitude too
large. On the other hand, with suspended
nanowires~\cite{husain03,feng07} or
nanotubes~\cite{yaish04,vanderzant06} with $d\approx 1$nm and
$m\approx10^{-21}$kg, also vibrating at $\Omega\approx10^8$Hz, we
obtain $m\Omega d^2\approx10^{-31}$ m$^2$kg/s, which is exactly what
we want.  Note that 100MHz is quite a high natural frequency for a
nanotube. For a lower frequency, one would obtain an even better value
for the effective $\hbar$, but this would require working at extremely
low temperatures, while for $\Omega\approx100$MHz we would need
$T_{env}\approx 10$mK. Thus, we believe that with slight improvement
over current technology, one would be able to see the
classical-to-quantum transition described here, by taking advantage of
the nonlinear nature of doubly-clamped nanoresonators.

%% If one looks at the resonator coupled to an environment, similar
%% behavior in the two distributions can be recognized, namely the
%% general outline of the Wigner function which resembles the classical
%% structure at short times, and the splitting of the distribution to two
%% parts centered around the two classically stable solutions at long
%% times. This is more evident for the higher temperature at which
%% $k_{B}T_{env}=2\hbar\Omega$, where the resonator is more classical.
%% In addition, the interference pattern in the Wigner function is
%% strongly suppressed by the coupling to an environment, after a time
%% usually referred to as `the decoherence time'.

Clearly, the most striking signature of quantum dynamics is the
appearance of quantum interference with negative regions in the Wigner
function. It would be wonderful if, at some point in the future, one
could directly measure the Wigner function and observe these negative
regions, as one can already do, for example, in the case of trapped
atoms~\cite{leibfried96}. Before such direct measurement is possible,
we suggest to perform experiments along the lines of the weak from of
QCT that we have studied here. In such experiments one should prepare
the system in a particular initial point $(x,p)$ in phase space, let it
evolve, and observe it at some later time $t$. Repeating the same
experiment 10,000 times, will allow one to generate a histogram of
measured positions $x(t)$, similar to the numerical histograms shown
in Fig.~\ref{nonlinear open kT0.2 hb0.1PD}. We emphasize that in this
kind of experimental protocol, the quantum state of the
system---assuming it is indeed quantum---is destroyed in the
measurement process, a time $t$ after it is initialized, but otherwise
one is free of any considerations of quantum measurement theory.

It is possible to recognize clear signatures while performing such an
experiment, implying that a driven Duffing resonator, working in its
bistability regime, is behaving in accordance with the laws of quantum
mechanics. In both cases one should operate in a regime of $\hbar$ and
$T_{env}$ similar to the one used here, in which thermal switching
between the two dynamical steady-states of the resonator is
exponentially suppressed, whereas quantum switching is possible. Under
such circumstances one can look for the following signatures:
\begin{enumerate}
  
\item If the initial coherent state is prepared within the basin of
  attraction of one of the two steady-state solutions of the dynamical
  system, as we did in section~\ref{open system T zero results}, the
  classical ensemble should flow as a whole to this solution at long
  times, while parts of the quantum Wigner function should flow to the
  other steady-state solution. This yields a finite probability of
  finding the quantum resonator oscillating in a dynamical state that
  is inaccessible for the classical resonator.
  
\item Even if the initial coherent state straddles over both basins of
  attraction, as we showed in section~\ref{open system T nonzero
    results}, and both distributions split between the two
  steady-state solutions, at long times there is a nonzero
  probability of finding the quantum resonator in a position between
  the two steady-state solutions, while classically this probability
  is zero.
\end{enumerate}

An interesting question that we have not yet addressed here, is
whether the quantum switching that we see takes place via tunneling or
quantum activation, as suggested by Marthaler and
Dykman~\cite{dykman06,dykman07}. We will address this question in a
future publication where we will consider individual quantum
trajectories of our system, possibly while performing a continuous
weak measurement, along the lines of the strong form of QCT.

\ack The authors are grateful to Michael Cross, Mark Dykman, Jens
Eisert, Victor Fleurov, Salman Habib, Inna Kozinsky, Michael Roukes,
and Keith Schwab for fruitful discussions. R.S. thanks the Lion
Foundation for supporting his stay at Tel Aviv University as an
exchange student. This research is supported by the U.S.-Isreal
Binational Science Foundation through grant No.~2004339, and by the
Israeli Ministry of Science.

\bigskip
\section*{References}
\bibliography{NJPpaper}

\begin{thebibliography}{10}

\bibitem{craighead00}
H.G. Craighead.
\newblock Nanoelectromechanical systems.
\newblock {\em Science}, 290:1532, 2000.

\bibitem{roukes01_02}
M.L. Roukes.
\newblock Plenty of room indeed.
\newblock {\em Scientific American}, 285:42, 2001.

\bibitem{roukes01_01}
M.L. Roukes.
\newblock Nanoelectromechanical systems face the future.
\newblock {\em Physics World}, 14:25, 2001.

\bibitem{C03}
A.N. Cleland.
\newblock {\em Foundations of Nanomechanics}.
\newblock Springer, Berlin, 2003.

\bibitem{EkinciRoukes05}
K.~L. Ekinci and M.~L. Roukes.
\newblock Nanoelectromechanical systems.
\newblock {\em Review of Scientific Instruments}, 76:061101, 2005.

\bibitem{cho03}
A.~Cho.
\newblock Researchers race to put the quantum into mechanics.
\newblock {\em Science}, 299:36--37, 2003.

\bibitem{blencowe04}
M.~P. Blencowe.
\newblock Quantum electromechanical systems.
\newblock {\em Physics Reports}, 395:159--222, 2004.

\bibitem{schwab_roukes05}
K.C. Schwab and M.L. Roukes.
\newblock Putting mechanics into quantum mechanics.
\newblock {\em Physics Today}, 58:36--42, July 2005.

\bibitem{knobel03}
R.G. Knobel and A.N. Cleland.
\newblock Nanometre-scale displacement sensing using a single electron
  transistor.
\newblock {\em Nature}, 424:291, 2003.

\bibitem{LaHaye04}
M.~D. LaHaye, O.~Buu, B.~Camarota, and K.~C. Schwab.
\newblock Approaching the quantum limit of a nanomechanical resonator.
\newblock {\em Science}, 304(5667):74--77, 2004.

\bibitem{Naik06}
A.~Naik, O.~Buu, M.~D. LaHaye, A.~D. Armour, A.~A. Clerk, M.~P. Blencowe, and
  K.~C. Schwab.
\newblock Cooling a nanomechanical resonator with quantum back-action.
\newblock {\em Nature}, 443(7108):193--196, 2006.

\bibitem{Penrose96}
Roger Penrose.
\newblock On gravity`s role in quantum state reduction.
\newblock {\em Gen.\ Relativ.\ Gravit.}, 28:581--600, 1996.

\bibitem{Leggett99}
A.~J. Leggett.
\newblock The significance of the {MQC} experiment.
\newblock {\em J.\ Supercond.}, 12(6):683--687, 1999.

\bibitem{Leggett02}
A.~J. Leggett.
\newblock Testing the limits of quantum mechanics: motivation, state of play,
  prospects.
\newblock {\em J.\ Phys.: Cond.\ Mat.}, 14(15):R415--R451, 2002.

\bibitem{cat}
E.~Schr\"{o}dinger.
\newblock {\em Naturwissenschaften}, 23:807, 1935.

\bibitem{HZMR03}
X.~M.~H. Huang, C.~A. Zorman, M.~Mehregany, and M.~L. Roukes.
\newblock Nanodevice motion at microwave frequencies.
\newblock {\em Nature}, 421:496, 2003.

\bibitem{cleland04}
A.~N. Cleland and M.~R. Geller.
\newblock Superconducting qubit storage and entanglement with nanomechanical
  resonators.
\newblock {\em Phys.\ Rev.\ Lett.}, 93:070501, 2004.

\bibitem{peano04}
V.~Peano and M.~Thorwart.
\newblock Macroscopic quantum effects in a strongly driven nanomechanical
  resonator.
\newblock {\em Physical Review B (Condensed Matter and Materials Physics)},
  70(23):235401, 2004.

\bibitem{Carr01}
S.~M. Carr, W.~E. Lawrence, and M.~N. Wybourne.
\newblock Accessibility of quantum effects in mesomechanical systems.
\newblock {\em Phys. Rev. B}, 64:220101(R), 2001.

\bibitem{Armour02}
A.~D. Armour, M.~P. Blencowe, and K.~C. Schwab.
\newblock Entanglement and decoherence of a micromechanical resonator via
  coupling to a cooper-pair box.
\newblock {\em Phys. Rev. Lett.}, 88:148301, 2002.

\bibitem{santamore04}
D.~H. Santamore, A.~C. Doherty, and M.~C. Cross.
\newblock Quantum nondemolition measurement of fock states of mesoscopic
  mechanical oscillators.
\newblock {\em Phys.\ Rev.\ B}, 70:144301, 2004.

\bibitem{Eisert04}
J.~Eisert, M.~B. Plenio, S.~Bose, and J.~Hartley.
\newblock Towards quantum entanglement in nanoelectromechanical devices.
\newblock {\em Phys.\ Rev.\ Lett.}, 93:190402, 2004.

\bibitem{myPRLarticle}
I.\ Katz, A.\ Retzker, R.\ Straub, and R.\ Lifshitz.
\newblock Signatures for a classical to quantum transition of a driven
  nonlinear nanomechanical resonator.
\newblock {\em Phys.\ Rev.\ Lett.}, 99(4):040404, 2007.

\bibitem{LCReview}
Ron Lifshitz and M.~C. Cross.
\newblock {\em Nonlinear dynamics of nanomechanical and micromechanical
  resonators}, volume~1 of {\em Review of Nonlinear Dynamics and Complexity},
  chapter~1.
\newblock Wiley, Berlin, 2008.

\bibitem{turner98}
K.~L. Turner, S.~A. Miller, P.~G. Hartwell, N.~C. MacDonald, S.~H. Strogatz,
  and S.~G. Adams.
\newblock Five parametric resonances in a microelectromechanical system.
\newblock {\em Nature}, 396:149--152, 1998.

\bibitem{BR00}
E.\ Buks and M.\~L.\ Roukes.
\newblock Metastability and the casimir effect in micromechanical systems.
\newblock {\em Europhys.\ Lett.}, 54:220, 2000.

\bibitem{BR02}
E.~Buks and M.~L. Roukes.
\newblock Electrically tunable collective response in a coupled micromechanical
  array.
\newblock {\em J. MEMS}, 11:802--807, 2002.

\bibitem{Blick02_2}
D.V. Scheible, A.\ Erbe, and R.H. Blick.
\newblock Evidence of a nanomechanical resonator being driven into chaotic
  response via the \uppercase{R}uelle-\uppercase{T}akens route.
\newblock {\em Appl.\ Phys.\ Lett.}, 81:1884, 2002.

\bibitem{yu02}
M.~Yu, G.~J. Wagner, R.~S. Ruoff, and M.~J. Dyer.
\newblock Realization of parametric resonances in a nanowire mechanical system
  with nanomanipulation inside a scanning electron microscope.
\newblock {\em Phys.\ Rev.\ B}, 66:073406, 2002.

\bibitem{aldridge05}
J.~S. Aldridge and A.~N. Cleland.
\newblock Noise-enabled precision measurements of a duffing nanomechanical
  resonator.
\newblock {\em Phys.\ Rev.\ Lett.}, 94(15):156403, 2005.

\bibitem{ina05}
H.W.Ch. Postma, I.~Kozinsky, A.~Husain, and M.L. Roukes.
\newblock Dynamic range of nanotube- and nanowire-based electromechanical
  systems.
\newblock {\em Appl.\ Phys.\ Lett.}, 86:223105, May 2005.

\bibitem{ina06}
I.\ Kozinsky, H.W.Ch. Postma, I.~Bargatin, and M.L. Roukes.
\newblock Tuning nonlinearity, dynamic range, and frequency of nanomechanical
  resonators.
\newblock {\em Appl.\ Phys.\ Lett.}, 88:253101, Jun 2006.

\bibitem{kozinsky07}
I.~Kozinsky, H.~W.~Ch. Postma, O.~Kogan, A.~Husain, and M.~L. Roukes.
\newblock Basins of attraction of a nonlinear nanomechanical resonator.
\newblock {\em Phys.\ Rev.\ Lett.}, 99:207201, 2007.

\bibitem{lifshitz03}
Ron Lifshitz and M.~C. Cross.
\newblock Response of parametrically driven nonlinear coupled oscillators with
  application to micromechanical and nanomechanical resonator arrays.
\newblock {\em Phys.\ Rev.\ B}, 67(13):134302, 2003.

\bibitem{bromberg06}
Yaron Bromberg, M.~C. Cross, and Ron Lifshitz.
\newblock Response of discrete nonlinear systems with many degrees of freedom.
\newblock {\em Phys.\ Rev.\ E}, 73(1):016214, 2006.

\bibitem{cross04}
M.~C. Cross, A.~Zumdieck, Ron Lifshitz, and J.~L. Rogers.
\newblock Synchronization by nonlinear frequency pulling.
\newblock {\em Phys. Rev. Lett.}, 93:224101, 2004.

\bibitem{cross06}
M.~C. Cross, J.~L. Rogers, Ron Lifshitz, and A.~Zumdieck.
\newblock Synchronization by reactive coupling and nonlinear frequency pulling.
\newblock {\em Phys. Rev. E}, 73(3):036205, 2006.

\bibitem{rugar91}
D.~Rugar and P.~Gr\"utter.
\newblock Mechanical parametric amplification and thermomechanical noise
  squeezing.
\newblock {\em Phys.\ Rev.\ Lett.}, 67(6):699--702, Aug 1991.

\bibitem{carr00}
D.~W. Carr, S.~Evoy, L.~Sekaric, H.~G. Craighead, and J.~M. Parpia.
\newblock Parametric amplification in a torsional microresonator.
\newblock {\em Appl. Phys. Lett.}, 77:1545--1547, 2000.

\bibitem{lupascu06}
A.~Lupascu, E.F.C. Driessen, L.~Roschier, C.J.P.M. Harmans, and J.E. Mooij.
\newblock High-contrast dispersive readout of a superconducting flux qubit
  using a nonlinear resonator.
\newblock {\em Phys.\ Rev.\ Lett.}, 96(12):127003, 2006.

\bibitem{milburn_schwab08}
M.J. Woolley, A.C. Doherty, G.J. Milburn, and K.C. Schwab.
\newblock Nanomechanical squeezing with detection via a microwave cavity.
\newblock {\em eprint arXiv:quant-ph/0803.1757v1}, 2008.

\bibitem{gardiner_zoller04}
C.W. Gardiner and P.~Zoller.
\newblock {\em Quantum Noise}.
\newblock Springer, Berlin, 3rd edition, 2004.

\bibitem{walls_milburn94}
D.F. Walls and G.J. Milburn.
\newblock {\em Quantum Optics}.
\newblock Springer, Berlin, 1994.

\bibitem{sergi-2007}
A.~Sergi and F.~Petruccione.
\newblock Nos\'e-{H}oover dynamics in quantum phase space.
\newblock {\em eprint arXiv:quant-ph/0711.2207v1}, 2007.

\bibitem{H&F}
Louis~N. Hand and Janet~D. Finch.
\newblock {\em Analytical Mechanics}.
\newblock Cambridge University Press, Cambridge, 1998.

\bibitem{feynman_hibbs}
R.~P. Feynman and A.~R. Hibbs.
\newblock {\em Quantum Mechanics and Path Integrals}.
\newblock McGraw-Hill, 1965.

\bibitem{schleich01}
W.P. Schleich.
\newblock {\em Quantum Optics in Phase Space}.
\newblock Wiley-VCH, Berlin, 2001.

\bibitem{berry89}
M.~V. Berry.
\newblock {\em Some Quantum-to-Classical Asymptotics}, volume LII of {\em Les
  Houches Lecture Series, Eds. M.J. Giannoni and A. Voros and J. Zinn-Justin},
  chapter~4.
\newblock Elsevier, 1991.

\bibitem{habibPRL}
S.~Habib, K.~Jacobs, H.~Mabuchi, R.~Ryne, K.~Shizume, and B.~Sundaram.
\newblock Quantum-classical transition in nonlinear dynamical systems.
\newblock {\em Phys.\ Rev.\ Lett.}, 88(4):040402, Jan 2002.

\bibitem{gat07}
Omri Gat.
\newblock Quantum dynamics and breakdown of classical realism in nonlinear
  oscillators.
\newblock {\em J.\ Phys.\ A:\ Math.\ Theor.}, 40:F911--F920, 2007.

\bibitem{berry78}
M.~V. Berry and N.~L. Balazs.
\newblock Evolution of semiclassical quantumm states in phase space.
\newblock {\em J. Phys. A: Math. Gen.}, 12(5):625--642, 1978.

\bibitem{habib98}
S.~Habib, K.~Shizume, and W.~H. Zurek.
\newblock Decoherence, chaos, and the correspondence principle.
\newblock {\em Phys. Rev. Lett.}, 80(20):4361--4365, May 1998.

\bibitem{cametti02}
F.~Cametti and C.~Presilla.
\newblock Quantum breaking time near classical equilibrium points.
\newblock {\em Phys.\ Rev.\ Lett.}, 89(4):040403, 2002.

\bibitem{oliveira03}
A.~C. Oliveira, M.~C. Nemes, and K.~M.~F. Romero.
\newblock Quantum time scales and the classical limit: Analytic results for
  some simple systems.
\newblock {\em Phys.\ Rev.\ E}, 68:036214, 2003.

\bibitem{oliveira06}
A.~C. Oliveira, J.~G.~Peixoto de~Faria, and M.~C. Nemes.
\newblock Quantum-classical transition of the open quartic oscillator: The role
  of the environment.
\newblock {\em Phys.\ Rev.\ E}, 73:046207, 2006.

\bibitem{Feynman_Vernon63}
R.~P. Feynman and F.~L. Vernon.
\newblock The theory of a general quantum system interacting with a linear
  dissipative system.
\newblock {\em Ann. Phys.}, 24:118--173, 1963.

\bibitem{caldeira_leggett83}
A.O. Caldeira and A.J. Leggett.
\newblock Quantum tunnelling in a dissipative system.
\newblock {\em Ann. Phys.}, 149(2):374--456, september 1983.

\bibitem{schlosshauer07}
Maximilian Schlosshauer.
\newblock {\em Decoherence and the quantum to classical transition}.
\newblock Springer, Berlin, 2007.

\bibitem{MCWF2}
M.B. Plenio and P.L. Knight.
\newblock The quantum-jump approach to dissipative dynamics in quantum optics.
\newblock {\em Rev.\ Mod.\ Phys.}, 70(1):101--144, Jan 1998.

\bibitem{greenbaum05}
B.~D. Greenbaum, S.~Habib, K.~Shizume, and B.~Sundaram.
\newblock The semiclassical regime of the chaotic quantum-classical transition.
\newblock {\em Chaos}, 15(3):033302, 2005.

\bibitem{habibLANL}
S.~Habib, T.~Bhattacharya, A.~Doherty, B.~Greenbaum, A.~Hopkins, K.~Jacobs,
  H.~Mabuchi, K.~Schwab, K.~Shizume, D.~Steck, and B.~Sundaram.
\newblock Nonlinear quantum dynamics.
\newblock {\em eprint arXiv:quant-ph/0505046}, 2005.

\bibitem{husain03}
A.~Husain, J.~Hone, Henk W.~Ch. Postma, X.~M.~H. Huang, T.~Drake, M.~Barbic,
  A.~Scherer, and M.~L. Roukes.
\newblock Nanowire-based very-high-frequency electromechanical resonator.
\newblock {\em Appl. Phys. Lett.}, 83(6):1240--1242, 2003.

\bibitem{feng07}
X.L. Feng, R.R. He, P.D. Yang, and M.L. Roukes.
\newblock Very high frequency silicon nanowire electromechanical resonators.
\newblock {\em Nano Lett.}, 7:1953--1959, Jul 2007.

\bibitem{yaish04}
V.~Sazonova, Y.\ Yaish, H.\ \"{U}st\"{u}nel, D.\ Roundy, T.A. Arias, and P.L.
  McEuan.
\newblock A tunable carbon nanotube electromechanical oscillator.
\newblock {\em Nature}, 431:284--287, Sep 2004.

\bibitem{vanderzant06}
B.\ Witkamp, M.\ Poot, and H.S.J. van~der Zant.
\newblock Bending-mode vibration of a suspended nanotube resonator.
\newblock {\em Nano Lett.}, 6:2904 -- 2908, Dec 2006.

\bibitem{louisell73}
W.~Louisell.
\newblock {\em Quantum Statistical Properties of Radiation}.
\newblock Wiley, New York, 1973.

\bibitem{spohn80}
H.~Spohn.
\newblock Kinetic equations from hamiltonian dynamics.
\newblock {\em Rev.\ Mod.\ Phys.}, 52:569--615, 1980.

\bibitem{MCWF}
K.~M{\o}lmer, Y.~Castin, and J.~Dalibard.
\newblock Monte carlo wave-function method in quantum optics.
\newblock {\em J. Opt. Soc. Am. B}, 10:524, 1993.

\bibitem{dykman06}
M.~Marthaler and M.~I. Dykman.
\newblock Switching via quantum activation: A parametrically modulated
  oscillator.
\newblock {\em Phys.\ Rev.\ A}, 73:042108, 2006.

\bibitem{dykman07}
M.~I. Dykman.
\newblock Critical exponents in metastable decay via quantum activation.
\newblock {\em Phys.\ Rev.\ E}, 75:011101, 2007.

\bibitem{leibfried96}
D.~Leibfried, D.M. Meekhof, B.E. King, C.~Monroe, W.M. Itano, and D.J.
  Wineland.
\newblock Experimental determination of the motional quantum state of a trapped
  atom.
\newblock {\em Phys.\ Rev.\ Lett.}, 77(21):4281--4285, Nov 1996.

\end{thebibliography}

\end{document}